\documentclass[preprint,number,sort&compress,twocolumn,3p]{elsarticle}
\pdfoutput=1
\usepackage{rotating}
\usepackage{graphicx}
\usepackage{array}
\usepackage{amsmath}
\usepackage{color}
\usepackage{amssymb}
\usepackage{slashed}
\usepackage[normalem]{ulem}
\usepackage[pdftex,hyperfootnotes=false,pdfpagelabels]{hyperref}

\begin{document}
  
\title{\mbox{Model-independent bounds on light pseudoscalars from rare B-meson decays}}

\tnotetext[t1]{This article is registered under preprint number: TTK-18-45}

\author[add1]{Babette D\"{o}brich}
\ead{babette.dobrich@cern.ch}
\author[add2]{Fatih Ertas}
\ead{ertas@physik.rwth-aachen.de}
\author[add2]{Felix Kahlhoefer}
\ead{kahlhoefer@physik.rwth-aachen.de}
\author[add3]{Tommaso Spadaro}
\ead{Tommaso.Spadaro@cern.ch}

\address[add1]{CERN,  1 Esplanade des Particules, CH-1211 Geneva 23, Switzerland}
\address[add2]{Institute for Theoretical Particle Physics and Cosmology (TTK), RWTH Aachen University, D-52056 Aachen, Germany}
\address[add3]{Frascati National Laboratory of INFN, Via E. Fermi, 40, 00044 Frascati (Roma), Italy}

\begin{abstract}
New light pseudoscalars, such as axion-like particles, appear in many well-motivated extensions of the Standard Model and provide an exciting target for present and future experiments. We study the experimental sensitivity for such particles by revising the CHARM exclusion contour, updating bounds from LHCb and presenting prospects for NA62 and SHiP. We first consider a simplified model of a light pseudoscalar $A$ and then propose a model-independent approach applicable to any spin-0 boson light enough to be produced in B-meson decays. As illustration, we provide upper bounds on $\text{BR}(B \to K\,A) \times \text{BR}(A \to \mu^+\mu^-)$ as a function of the boson lifetime and mass for models that satisfy minimal flavour violation. Our results demonstrate the important complementarity between different experiments resulting from their different geometries.

\bigskip 

\end{abstract}


\maketitle

\section{Introduction}

The absence of evidence for new physics at the TeV scale so far has led to a renewed attention for much lighter new particles, which may have evaded detection due to their very weak interactions with Standard Model (SM) particles. One of the most attractive realisations of this idea are light pseudoscalars, often called axion-like particles (ALPs), for which the smallness of the mass and the interactions can simultaneously be explained if they arise as Pseudo-Goldstone bosons of a spontaneously broken approximate global symmetry. These particles may couple to SM particles via the so-called axion portal~\cite{Nomura:2008ru,Batell:2009di,Freytsis:2009ct,Hochberg:2018rjs} and can be searched for in a wide range of experiments both at the high-energy frontier of particle physics~\cite{Mimasu:2014nea,Jaeckel:2015jla,Brivio:2017ije,Bauer:2017ris,Haisch:2018kqx} and at the so-called intensity frontier~\cite{Batell:2009di,Andreas:2010ms,Hewett:2012ns,Essig:2013lka,Flacke:2016szy}.

Light pseudoscalars have received particular attention in the context of dark matter model building. The reason is that if the interactions between dark matter and SM particles are mediated by a pseudoscalar exchange particle, one expects a strong suppression of event rates in direct detection experiments, consistent with the non-observation of a dark matter signal in these experiments~\cite{Nomura:2008ru,Freytsis:2010ne,Dienes:2013xya,Arina:2014yna}.\footnote{It was recently pointed out that scattering in direct detection experiments may in fact be dominated by loop processes that are unsuppressed in the non-relativistic limit~\cite{Arcadi:2017wqi,Sanderson:2018lmj,Abe:2018emu}. But even when including these effects the expected event rates are still so small that a detection will be very challenging.} There has consequently been substantial effort to construct and study models with pseudoscalar mediators in the context of dark matter indirect detection~\cite{Boehm:2014hva,Ipek:2014gua,Berlin:2015wwa,Tunney:2017yfp}, collider searches~\cite{No:2015xqa,Goncalves:2016iyg,Bauer:2017ota} and self-interacting dark matter~\cite{Kahlhoefer:2017umn,Hochberg:2018rjs}.

As pointed out in Refs.~\cite{Freytsis:2009ct,Batell:2009jf,Andreas:2010ms,Dolan:2014ska}, models with light pseudoscalars face strong constraints from rare decays. The reason is that any pseudoscalar coupling to SM quarks will at the one-loop level induce flavour-changing processes such as $b\to s A$ or $s \to d A$, which can be searched for with great sensitivity in a number of experiments~\cite{Hiller:2004ii}. In the mass range where these constraints are relevant ($m_A \lesssim 5\,\mathrm{GeV}$) it is therefore very difficult to obtain sizeable interactions between dark matter and SM quarks via the exchange of a pseudoscalar mediator. Nevertheless, Ref.~\cite{Dolan:2014ska} also found that certain parameter regions are rather difficult to probe experimentally. For example, pseudoscalars with masses between a few hundred MeV and a few GeV and lifetimes of the order of $10^{-9}\;\mathrm{s}$ are very difficult to constrain, since the resulting decay length is of the order of meters, which is large compared to the scales of colliders but short on the scale of fixed-target experiments. It is therefore an interesting and important question to understand which experiments are capable of probing this parameter region. Among the possible contenders are a number of existing and planned fixed-target experiments with relatively short absorber length, such as NA62~\cite{Hahn:1404985,Dobrich:2015jyk}, SeaQuest~\cite{Aidala:2017ofy,Berlin:2018pwi} or SHiP~\cite{Alekhin:2015byh,Anelli:2015pba}, collider experiments with sensitivity to displaced vertices, such as LHCb using the vertex locator VELO~\cite{LHCbVELOGroup:2014uea}, or dedicated forward detectors like FASER~\cite{Feng:2017uoz,Feng:2018noy}.

In the present work, we address this question by improving upon the analysis presented in Ref.~\cite{Dolan:2014ska} in four key regards. First, we perform a more careful calculation of constraints and projected sensitivities from fixed-target experiments, taking into account the detailed energy distribution of B-mesons produced in the target as well as the propagation of the resulting particles and the detector geometry. Second, we include new constraints from searches for $\mu^+\mu^-$ pairs originating from a displaced vertex in rare B-meson decays at LHCb~\cite{Aaij:2015tna,Aaij:2016qsm}. Third, we include updated calculations of the branching ratios of light pseudoscalars from Ref.~\cite{Domingo:2016yih}. Finally, we develop a new approach for presenting experimental bounds that does not require any assumptions on the underlying theory. The aim of all of these improvements is to study in a model-independent way which regions or parameter space remain unexplored and whether they can be probed with future fixed-target experiments.

This work is structured as follows. In Section~\ref{sec:model} we introduce the effective interactions that we want to investigate and present the relevant expressions for the partial decay widths of rare B-meson decays and pseudoscalar decays. Section~\ref{sec:experiments} then concerns the calculation of updated experimental constraints and projections. Our new approach for presenting experimental bounds will be discussed in Section~\ref{sec:model-independent}, followed by the conclusions in Section~\ref{sec:conclusions}.

\section{Light pseudoscalars and rare decays}
\label{sec:model}

We consider a light pseudoscalar $A$ with Yukawa-like couplings to SM fermions:
\begin{equation}
\mathcal{L} = i  \, g_Y \sum_{f = q,\ell} \frac{m_f}{v} A \, \bar{f} \gamma^5 f \; ,
\label{eq:L}
\end{equation}
where $m_f$ is the fermion mass, $v\simeq246\:\text{GeV}$ is the electroweak vacuum expectation value and $q = \{u, d, s, c, b, t\}$ and $\ell = \{e, \mu, \tau\}$ denote SM quarks and charged leptons, respectively. As pointed out by a number of recent studies~\cite{Ipek:2014gua,No:2015xqa,Goncalves:2016iyg,Bauer:2017ota}, the Lagrangian in eq.~(\ref{eq:L}) does not respect the gauge symmetries of the full unbroken SM gauge group and should therefore be thought of as the low-energy limit of a more complete theory, such as a Two-Higgs Doublet Model (2HDM).

At the one-loop level this Lagrangian induces flavour-changing neutral currents. For the present work we are particularly interested in the one leading to the transition $b \to s A$, which can be written as:
\begin{equation}
\mathcal{L}_\text{FCNC} \supset h^R_{sb} A \, \bar{s}_L b_R + h^L_{sb} A \, \bar{s}_R b_L 
+ \text{h.c.}
\label{eq:hRhL}
\end{equation}
with $q_{R,L} = \frac{1}{2}(1\pm\gamma^5) q$. As expected for an effective theory, the relevant loop diagrams are divergent, such that the coefficients can be parametrised in terms of a new-physics scale $\Lambda$~\cite{Dolan:2014ska}:
\begin{equation}
 h^R_{sb} = \frac{i \alpha \, g_Y \, m_b \, m_t^2}{4 \pi \, m_W^2 \, \sin(\theta_W)^2 \, v} \, V_{tb} V_{ts}^\ast \, \log\left(\frac{\Lambda^2}{m_t^2}\right) \; ,
 \label{eq:FCNC}
\end{equation}
where $m_W$ is the $W$-boson mass, $\alpha \equiv e^2 / (4 \pi)$, $\theta_W$ is the Weinberg angle and $V$ is the CKM matrix. The corresponding expression for $h^L_{sb}$ (up to an overall sign) is obtained by replacing $m_b$ by $m_s$.

Defining
\begin{equation}
h^S_{sb} = (h^R_{sb} + h^L_{sb})/2\;, \qquad h^P_{sb} = (h^R_{sb} - h^L_{sb})/2 \; ,
\label{eq:hShP}
\end{equation}
the partial decay widths for the corresponding flavour-changing B-meson decays are given by~\cite{Bobeth:2001sq, Hiller:2004ii, Batell:2009jf,Choi:2017gpf}:
\begin{align}
\Gamma(B \rightarrow K \, A) = & \frac{|h^S_{sb}|^2 }{16 \pi \, m_{B}^3} \lambda^{1/2}(m_{B}^2, m_{K}^2, m_A^2) \nonumber \\ & \times \left| f_0^{B^0}(m_A^2)\right|^2 \left(\frac{m_{B}^2 - m_{K}^2}{m_b - m_s}\right)^2 \; , \\
\Gamma(B \rightarrow K^\ast \, A) = & \frac{|h^P_{sb}|^2  }{16 \pi \, m_{B}^3} \lambda^{3/2}(m_{B}^2, m_{K^\ast}^2, m_A^2) \nonumber \\ & \times \left| A_0^{B^0}(m_A^2)\right|^2 \frac{1}{\left(m_b + m_s\right)^2}\; ,
\label{eq:Bwidth}
\end{align}
with $\lambda(a,b,c) =(a-b-c)^2-4\,b\,c$. The relevant form factors are given by~\cite{Ali:1999mm,Ball:2004ye,Ball:2004rg}
\begin{align}
f_0^{B^0}(q^2) & = \frac{0.33}{1-\tfrac{q^2}{38\:\text{GeV}^2}} \; , \\
A_0^{B^0}(q^2) & = \frac{1.36}{1-\tfrac{q^2}{28\:\text{GeV}^2}} - \frac{0.99}{1-\tfrac{q^2}{37\:\text{GeV}^2}} \; .
\end{align}
D-meson decays involving $A$, on the other hand, suffer from CKM as well as mass suppression in eq.~(\ref{eq:FCNC}) and are therefore not taken into consideration~\cite{Bezrukov:2009yw}.

Finally, we need to consider the decay of $A$ into SM particles. If the pseudoscalar has no interactions apart from the ones given in eq.~(\ref{eq:L}), the dominant decay modes are expected to be $A \to \ell^+ \ell^-$ for most of the mass range that we will be interested in.\footnote{In particular, we assume that the pseudoscalar has no enhanced coupling to photons. 
For a detailed discussion of constraints on ALPs that couple dominantly to photons, we refer to~\cite{Dobrich:2015jyk,Dolan:2017osp,Arias-Aragon:2017eww}.} The corresponding partial decay width is given by
\begin{equation}
\Gamma(A \rightarrow \ell^+ \ell^-) = \frac{g_Y^2 \, m_\ell^2}{8 \pi \, v^2} \, m_A \, \sqrt{1 - \frac{1}{z_\ell}} \, ,\\
\end{equation}
where $z_\ell = m_A^2 / (4 \, m_\ell^2)$. The precise branching ratios for light pseudoscalars are however difficult to calculate due to large uncertainties in the partial width for hadronic decay modes~\cite{Bauer:2017ris}. Here we adopt the recent results from Ref.~\cite{Domingo:2016yih}, which estimates the hadronic decay width using Chiral Perturbation Theory for $m_A < 1.2\,\mathrm{GeV}$ and a spectator quark model for $1.2\,\mathrm{GeV} < m_A < 3\,\mathrm{GeV}$.\footnote{Note that, although these results have been obtained in the context of the NMSSM, they can directly be applied to our case. In particular, since the most important contribution to the hadronic decay width stems from the pseudoscalar coupling to strange quarks, the leptonic branching ratios are to first approximation independent of $\tan \beta$.} For $m_A > 3\,\mathrm{GeV}$ we extrapolate the results from Ref.~\cite{Domingo:2016yih} by assuming that the hadronic decay width converges to the partonic width for a pair of free strange quarks with mass $m_s = 95\,\mathrm{MeV}$ and that decays involving charm quarks can be neglected for $m_A < 4.7\,\mathrm{GeV}$. We will return to the question of how to deal with the uncertainties related to the muonic branching ratio in Section~\ref{sec:model-independent}.

\section{Experimental constraints}
\label{sec:experiments}

In the following we will focus on pseudoscalars that decay predominantly into pairs of muons and taus, i.e.\ $m_A \gtrsim 210\,\mathrm{MeV}$, but are light enough to be produced in B-meson decays, $m_A \lesssim 4.7\,\mathrm{GeV}$. The dominant constraints in this mass range come from LHCb, which is most sensitive for relatively short-lived pseudoscalars, and fixed-target experiments, which can probe pseudoscalars with very small couplings and rather long lifetime.

\subsection{Fixed-target experiments}

\begin{figure}[t]
\begin{center}
\includegraphics[width=7cm]{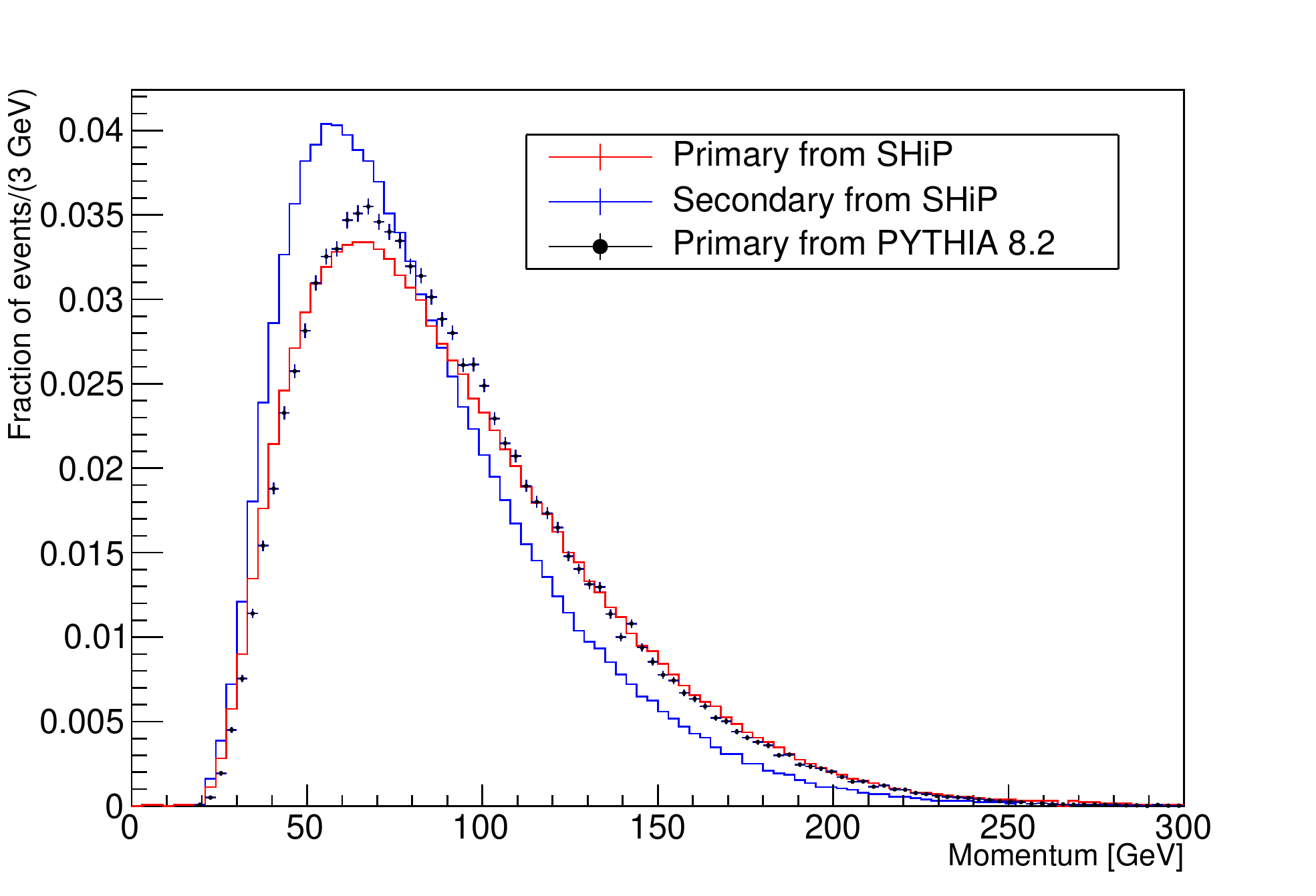}

\includegraphics[width=7cm]{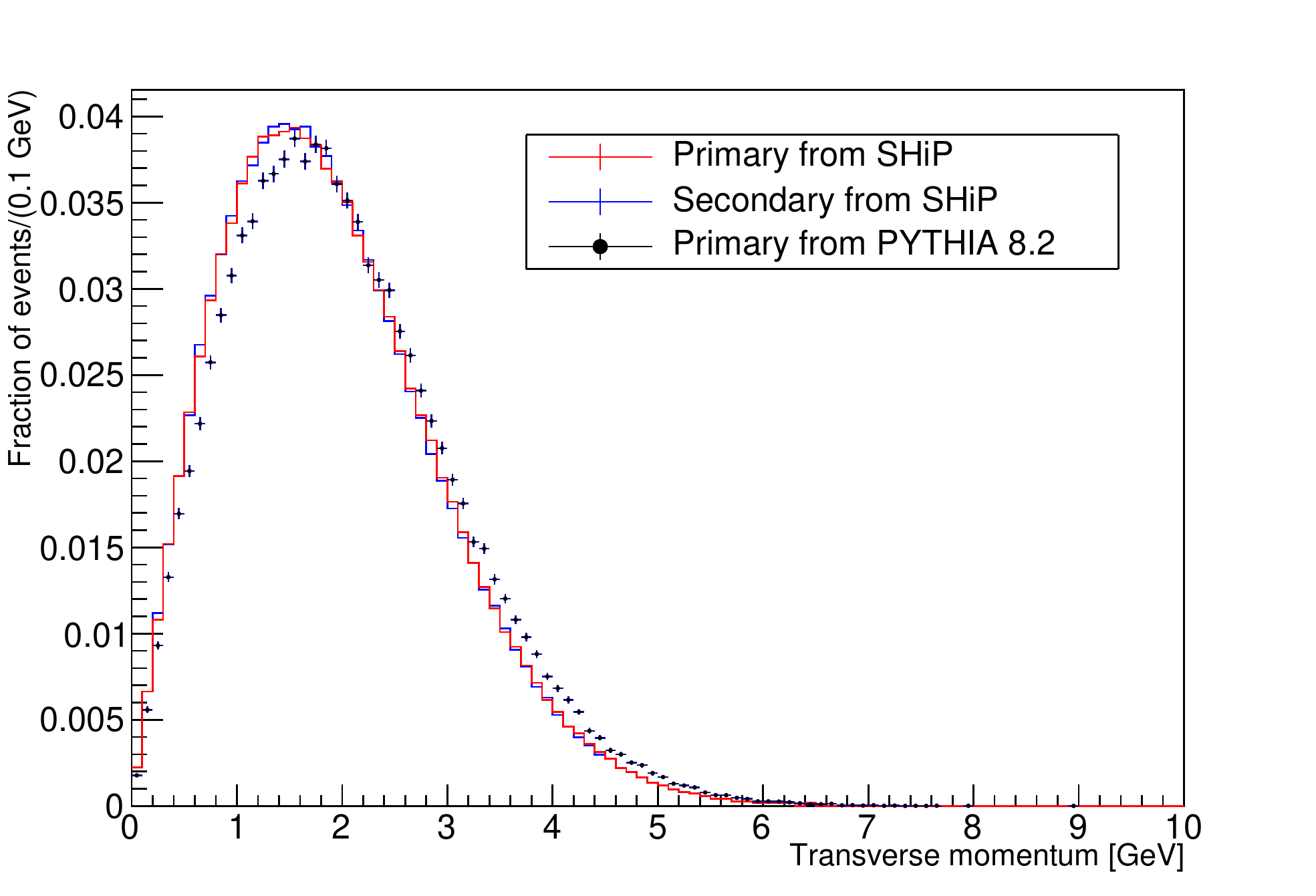}
\caption{\label{fig:ship_overlay}
Top (bottom): comparison of momentum (transverse momentum) distributions for $B^0$, $B^+$, $B_{S}$ and the corresponding anti-particles produced by a 400-GeV proton beam interacting onto a fixed proton target. 
Vertical axes: fractional yield. 
Dark blue line: distribution from \cite{CERN-SHiP-NOTE-2015-009,Lanfranchi:2243034}, for secondary production.
Red line: distribution used from \cite{CERN-SHiP-NOTE-2015-009,Lanfranchi:2243034}, for primary yield.
Solid dots: production used for the estimate in our toy MC, from \texttt{PYTHIA} version 8.2.}
 \end{center}
 \end{figure}
 
To calculate constraints on light pseudoscalars from fixed-target experiments, the first challenge is the production of B-meson spectra. To this purpose we employ \texttt{PYTHIA 8.23}~\cite{Sjostrand:2014zea}, solely allowing the simulation of hard-QCD bottom-quark production processes. The PDF set number 2, corresponding to the leading-order 5L set from the CTEQ collaboration, has been used~\cite{Nadolsky:2008zw}. Different evaluations have been made while varying the minimum values of the invariant transverse momentum and of the invariant mass. The chosen values, 300~MeV for both of them, have been optimised after comparing the cross section evaluated from \texttt{PYTHIA} with the experimental value 
extrapolated to an energy of 27.4~GeV in the center of mass, corresponding to a 400 GeV beam. The $pp\to b\bar{b}$ cross section evaluated from \texttt{PYTHIA} is 1.866(3)~nb, in good agreement with the experimental data shown in figure~30 of~\cite{Lourenco:2006vw}. We have cross-checked our spectra against the primary yield computed for SHiP~\cite{CERN-SHiP-NOTE-2015-009} and include with our work an auxiliary file with the spectra.\footnote{The file ``B400GeV.root'' contains a \texttt{ROOT Tree} of the momenta of $B^0$, $B_S^0$, and $B^+$ and their anti-particles, produced after 400~GeV proton beam interactions onto a fixed proton target. The $z$ component is along the beam line.} Figure \ref{fig:ship_overlay} shows a comparison of the B-meson spectra used in our simulations with the ones found in Refs.~\cite{CERN-SHiP-NOTE-2015-009,Lanfranchi:2243034}. As outlined in Ref.~\cite{CERN-SHiP-NOTE-2015-009}, in principle it would be possible to also include the yield of B-mesons from secondary proton interaction, which can give a relevant contribution to the overall yield. This contribution is however not included in our study, as it would require much more computationally expensive simulations.

To translate the yield obtained from \texttt{PYTHIA} into a total yield, one has to weight the Monte Carlo (MC) events with the ratio of the $bb$ cross section and the total proton-proton cross section at $400\,\mathrm{GeV}$. The total proton-proton cross section is taken
to be $\sigma(pp)\simeq 40\,\mathrm{mb}$, see e.g.\ Ref.~\cite{CERN-SHiP-NOTE-2015-009} (figure 1). The dependence on the target material is accounted for by a correction factor of $A^{1/3}$~\cite{Lourenco:2006vw}.

In this study we consider existing bounds from the past CHARM neutrino experiment and prospects for the running NA62 experiment as well as the proposed SHiP facility.\footnote{Another fixed-target experiment of interest is SeaQuest~\cite{Aidala:2017ofy,Berlin:2018pwi}, which operates with a beam energy of $120\,\mathrm{GeV}$. However, to the best of our knowledge, there are no measurements of the $bb$ cross section at a centre-of-mass energy of $15.1\,\mathrm{GeV}$, making it impossible to validate the $bb$ cross section calculated by \texttt{PYTHIA}. We therefore do not include SeaQuest in our analysis.} 
To estimate the experimental sensitivities we rely on a toy MC with geometries as outlined below, do not account for efficiencies (unless
stated otherwise) and assume that backgrounds are negligible. Our results should therefore not be interpreted as an accurate prediction of what may ultimately be achievable in these experiments after accounting for selection efficiencies and potential backgrounds.

The toy MC for all fixed-target set-ups under consideration has the following structure: First we randomly sample B-mesons ($B^0$ or $B^+$ and their anti-particles) from the \texttt{PYTHIA} output. Then we randomly sample from the decays $B \rightarrow K + A$ and $B \rightarrow K^{\star} + A$ by summing their respective decay widths and accounting for their relative probability. Finally, we calculate the probability that the pseudoscalar decays inside the decay volume of the experiment:
\begin{equation}
p(l_A) = \exp\left(-\frac{d}{l_A}\right) - \exp\left(-\frac{d+l}{l_A}\right) \, ,
\end{equation}
where $l_A = E_A \tau_A / m_A$ is the pseudoscalar decay length, $d$ denotes the distance of the pseudoscalar production point to the decay volume and $l$ is the fiducial length in which pseudoscalar decays can be detected.

For CHARM~\cite{Bergsma:1985qz,Bergsma:1983rt}, the following input enters our toy MC: The detector was located at the comparably far distance $d=480\,\mathrm{m}$ away from the proton dump. The detector (here the same as the fiducial volume) was $l=35\,\mathrm{m}$ in length and $3\times3\,\mathrm{m}$ 
in transverse dimensions. The transversal offset of $5\,\mathrm{m}$ from the beam axis is accounted for in the MC. According to Ref.~\cite{Bergsma:1985qz}, CHARM was sensitive to events with {\it one or two} detected muons in acceptance. Considering CHARM's material budget in front of the downstream calorimeter we require at least one muon of energy greater than $2\,\mathrm{GeV}$ in the detector acceptance for the event to be accepted with full efficiency. CHARM quotes a signal acceptance of 85\%, which we account for in our estimate. The target material is copper and the number of protons on target (POT) is $2.4 \times 10^{18}$.

The NA62 experiment~\cite{NA62:2017rwk} aims at a precise measurement of the ultra-rare decay $K^+\rightarrow \pi^+ \nu \bar{\nu}$. Besides this main measurement, NA62 has a rich programme for exotic particles, including long-lived particles that can be produced in the up-stream copper beam collimator, where the primary SPS proton-beam is fully (in dump-mode) or partially (in standard data-taking) dumped. Here we consider proton interactions that yield B-mesons in the copper-beam collimator located $\simeq 23\,\mathrm{m}$ downstream the primary target.\footnote{The `$0\,\mathrm{m}$' position at NA62 is the location of a Beryllium target, which is the source of the secondary kaons used for NA62's main measurement. Since NA62 is a `kaon factory', the consideration of pseudoscalar production in upstream or downstream kaon decays is an interesting topic that goes beyond the scope of this work.} 
We model NA62 in our MC according to the following parameters: The distance between the beam-defining collimator and the start of the fiducial volume is $d = 82\,\mathrm{m}$ and the vacuum decay region before the spectrometer is $l=75\,\mathrm{m}$ long. In addition, we require the following acceptance conditions: Both muons produced in the pseudoscalar decay need to be detected at the first and last spectrometer chamber as well as at the calorimeter (for all of these, the central hole that allows the beam to pass has been accounted for). Also, we require that each $\mu$ has an energy greater than $5\,\mathrm{GeV}$ (in order to be efficiently tracked by the trackers with the standard pattern recognition). Finally, each $\mu$, when extrapolated to the `MUV3-scintillator' plane, is within its acceptance. The target material for the production of B-mesons is copper and we show NA62 prospects for $1 \times 10^{18}$ POT.

Finally, we model the prospects for SHiP~\cite{Alekhin:2015byh,Anelli:2015pba} as follows. We implement a spectrometer with the following geometrical parameters~\cite{SHiPSpectrometer}: The fiducial region is taken to be $45\,\mathrm{m}$ downstream from the production point. The spectrometer magnet is positioned at $32.5\,\mathrm{m}$ behind the start of the fiducial region and gives a kick with $0.75\,\mathrm{T\,m}$. As a minimal requirement, we ask both $\mu$ tracks to be within two spectrometer chambers located at $28\,\mathrm{m}$ and $37\,\mathrm{m}$ behind the start of the fiducial region, respectively. The spectrometer chambers are  ellipsoidal regions with half-axes $x=2.5\,\mathrm{m}$ and $y=5\,\mathrm{m}$. In addition, both muons need to be detected at the end of a fiducial region of $50\,\mathrm{m}$ in the same ellipsoid and we ask for the total energy in acceptance to be greater than $2\,\mathrm{GeV}$. The target material for the production of B-mesons is taken to be molybdenum and we show SHiP prospects for $1 \times 10^{20}$ POT.

\subsection{LHCb}

LHCb has performed searches for displaced vertices in the processes $B \to K \phi$ and $B \to K^\ast \phi$, where $\phi$ is a scalar boson that subsequently decays into a di-muon pair~\cite{Aaij:2015tna,Aaij:2016qsm}. Since both the B-meson and the scalar boson $\phi$ decay isotropically, these searches can be directly reinterpreted for the case of a pseudoscalar instead of a scalar. The results from LHCb then provide upper bounds on $\text{BR}(B \to K A)\times\text{BR}(A \to \mu^+\mu^-)$ and $\text{BR}(B \to K^\ast A)\times\text{BR}(A \to \mu^+\mu^-)$ as a function of the pseudoscalar mass $m_A$ and its lifetime $\tau_A$.

For $B \to K^\ast \phi$ LHCb provides a numerical code with an interpolation of the experimental results~\cite{LHCbcode}. For given $m_A$ and $g_Y$ we therefore simply need to calculate the model prediction for $\text{BR}(B \to K^\ast A)\times\text{BR}(A \to \mu^+\mu^-)$ and the lifetime $\tau_A$ and then apply the code to determine whether the point is excluded by LHCb data. For $B \to K \phi$, we make use of a digitised version of figure~4 from Ref.~\cite{Aaij:2016qsm} to implement a similar approach. 

Since events with a di-muon invariant mass in the $J/\psi$, $\psi(2S)$ and $\psi(3770)$ resonance regions are not considered in the LHCb analyses, we do not obtain any constraints for pseudoscalar masses close to these resonances.\footnote{In the analysis of $B\to K \phi$, LHCb has also fully vetoed the $K^0_S$ resonance. Moreover, the $\psi(4160)$ resonance is also excluded in the digitised data that we have used. However, in both of these regions we obtain bounds from the $B \to K^\ast \phi$ data.} The strongest constraints in these regions therefore still come from the analysis in Ref.~\cite{Dolan:2014ska} of prompt decays in LHCb~\cite{Aaij:2012vr}.

\subsection{Results}

\begin{figure}[t]
  \centering
  \includegraphics[width=0.4\textwidth]{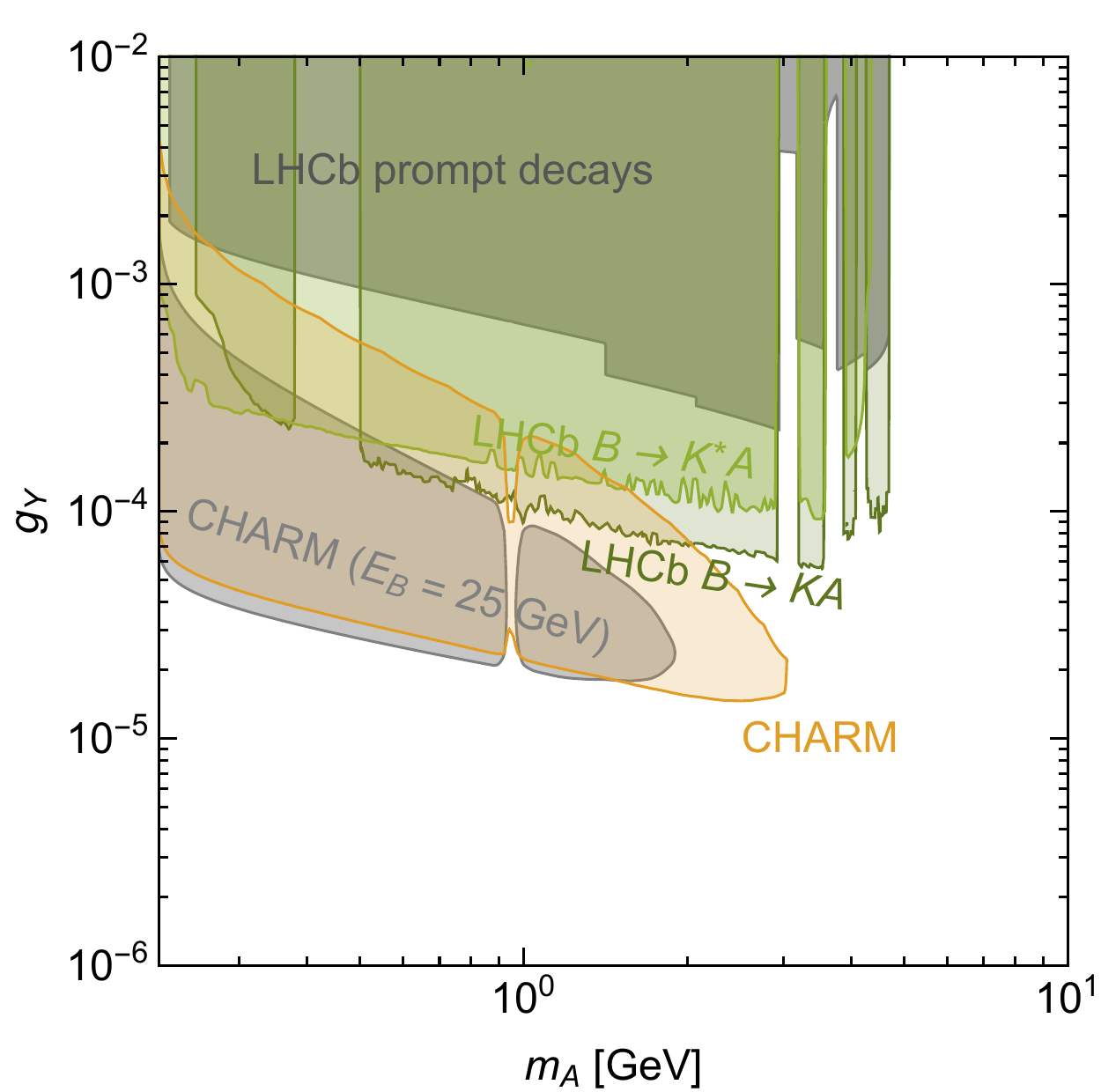}\qquad
  
  \includegraphics[width=0.4\textwidth]{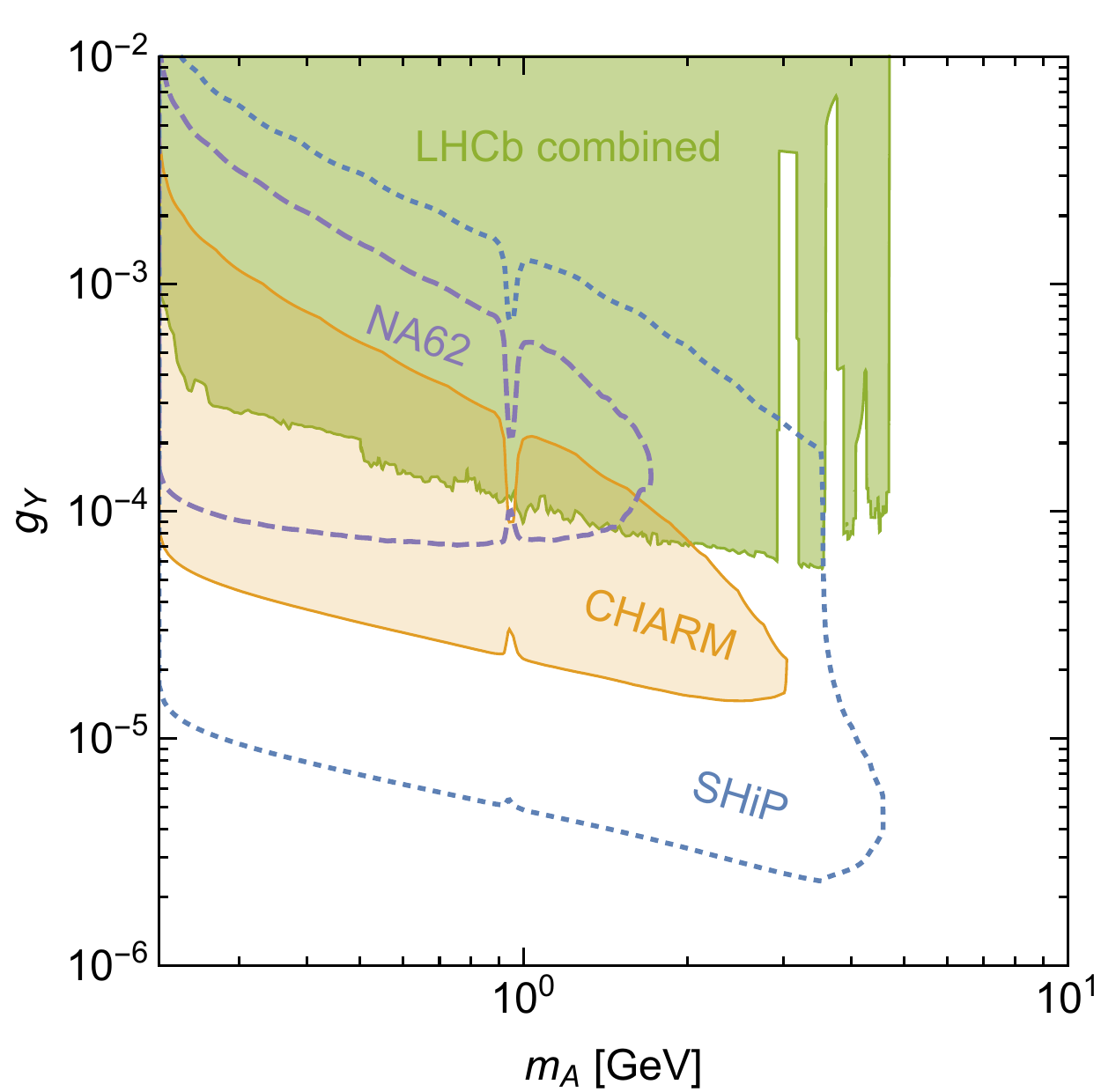}
  \caption{Constraints on light pseudoscalars from LHCb and CHARM at 95\% confidence level compared to previous results from the literature (top) and compared to the projected sensitivities of NA62 and SHiP (bottom).}
  \label{fig:results}
\end{figure}

The results of our analysis are summarised in figure~\ref{fig:results}. In the upper panel we compare the new exclusion limits (coloured regions) to the ones obtained in previous analyses of light pseudoscalars (grey), which only included prompt decays for LHCb and assumed a fixed B-meson energy for CHARM~\cite{Dolan:2014ska}.\footnote{Note that the definition of $g_Y$ adopted in the present work differs by a factor of $\sqrt{2}$ from the one in Ref.~\cite{Dolan:2014ska}. Also, the CHARM bound shown in Ref.~\cite{Dolan:2014ska} included a contribution from the process $K \to \pi A$, even though kaons loose a substantial fraction of their energy in the target before decaying and therefore do not produce a collimated beam of pseudoscalars. 
The grey region shown in figure~\ref{fig:results}, in contrast, includes only pseudoscalars produced via $B \to K A$ and $B \to K^\ast A$.} The shift of the updated CHARM contour towards larger couplings is a direct consequence of the improved calculation of the B-meson energy spectrum. In fact, we find that the B-meson energy for pseudoscalars in the signal region peaks significantly  above $25\,\mathrm{GeV}$, depending on the value of $m_A$. The increased boost factor means that pseudoscalars with larger couplings (and hence shorter lifetime) can still induce a signal in the detector behind the absorber.

Conversely, by including searches for displaced vertices the exclusion limits from LHCb are extended towards significantly longer pseudoscalar lifetimes, i.e.\ smaller couplings. We find that the improved treatment of CHARM in combination with the updated constraints from LHCb almost completely closes the gap between the two experiments that previously existed around couplings of order $10^{-4}$. Indeed, for $2 m_\mu < m_A \lesssim 3\,\mathrm{GeV}$ the combined constraints imply $g_Y \lesssim 5\times10^{-5}$.

We also observe a suppression of experimental sensitivies for $m_A \approx 958\,\mathrm{MeV}$. This suppression is the result of mixing between the new pseudoscalar and the $\eta'$-meson, which substantially increases the hadronic decay width of the pseudoscalar. As a result, the muonic branching ratio is suppressed and the pseudoscalar lifetime is reduced, which in combination leads to a suppression of the expected number of observable decays.\footnote{If we were to include hadronic final states in our analysis, experimental sensitivity would actually \emph{increase} for $m_A \approx m_{\eta'}$ and very small couplings. Lacking an appropriate MC simulation for hadronic final states, we include only muons and thus provide a conservative  sensitivity estimate. Note also that we do not include additional contributions to the pseudoscalar production rate resulting from mixing (see also Section~\ref{sec:model-independent}).} A similar effect occurs for $m_A \approx m_\eta = 548\,\mathrm{MeV}$, but the width of the resulting dip is too narrow to be resolved.

In the lower panel we compare the existing constraints from CHARM and LHCb with the projected sensitivities of NA62 and SHiP. As expected, SHiP promises a substantial gain in sensitivity compared to existing experiments.\footnote{In analogy to the improvement for CHARM, we find that the expected sensitivity of SHiP significantly exceeds the one previously published in Ref.~\cite{Alekhin:2015byh}, where once again a fixed B-meson energy was assumed.} On the other hand, NA62 on first sight appears not to provide any new information, as the parameter region that can be probed falls exactly into the region excluded by the improved analyses of CHARM and LHCb. Nevertheless, we emphasise that this conclusion only holds in the context of the specific model that we have considered so far. In the following section, we will show by adopting a model-independent approach that NA62 can in fact probe parameter regions excluded by neither CHARM nor LHCb.

\section{Model-independent bounds}
\label{sec:model-independent}

The results presented in the previous section have been obtained under a number of assumptions. First of all, we calculated the total decay width of $A$ by adopting a specific calculation of the hadronic decay width and neglecting final states involving charm quarks. The corresponding uncertainties, in particular concerning the mixing between $A$ and pseudoscalar mesons, are however difficult to quantify. Moreover, it is conceivable that the pseudoscalar couples with different strength to quarks and leptons, for example in lepton-specific 2HDMs~\cite{Branco:2011iw}, which would modify our results. An easy way to address these issues and to allow for a broader reinterpretation of our results would be to quote the largest value of $\text{BR}(A\to\mu^+\mu^-)$ that can be excluded experimentally for each point in $g_Y$--$m_A$ parameter space.

We have, however, also made another more troublesome assumption by fixing the relation between the production rate of $A$ and its subsequent lifetime via eq.~(\ref{eq:FCNC}) with a specific choice of $\Lambda$. In fact, these two quantities enter into the calculation of experimental bounds in very different ways: The expected number of events in a given experiment typically depends linearly on the production rate (i.e.\ the B-meson branching ratios) but exponentially on the pseudoscalar lifetime. It is thus impossible to use the results presented above to infer the corresponding constraints for a different value of $\Lambda$ using a simple rescaling.

At first sight, this problem is not very severe, given that $\Lambda$ only enters logarithmically. However, it has been shown that the flavour-changing interactions can in fact vary quite substantially between different high-energy theories that lead to the same effective interaction between pseudoscalars and quarks~\cite{Choi:2017gpf}. For example, an important contribution to flavour-changing processes may arise from pseudoscalar-Higgs interactions or interactions between the pseudoscalar and $SU(2)_L$ gauge bosons~\cite{Izaguirre:2016dfi}. Furthermore, in specific UV completions there may be additional non-logarithmic contributions, which become important if $\Lambda$ is small, in particular if they have the opposite sign~\cite{Ellwanger:2016wfe}. Finally, mixing between the pseudoscalar and QCD resonances may enhance the rate of rare B-meson decays for specific pseudoscalar masses.

It therefore becomes an important problem to present results in such a way that they do not rely on a specific relation between production and decay. Indeed, such model-independent approaches are frequently employed in the context of searches for long-lived particles (see e.g.\ Refs.~\cite{Aaboud:2018jbr,Sirunyan:2018pwn}). The crucial idea is to treat the production mechanism and the decay length as fully independent parameters, rather than attempting to derive them from the same fundamental interactions. In our case, this means that rather than presenting experimental bounds as a function of $g_\mathrm{Y}$ and $m_A$, we present them as a function of $\mu_A \equiv \text{BR}(B \to K A) \times \text{BR}(A \to \mu^+ \mu^-)$, $\tau_A$ and $m_A$. The fact that we now have three rather than two independent parameters is a direct consequence of the need to capture a broader range of possible underlying theories.

For a single experiment, the most convenient way to present experimental bounds is to provide upper bounds on $\mu_A$ as a function of $\tau_A$ and $m_A$, as done for example in~\cite{Aaij:2015tna,Aaij:2016qsm}. This approach makes it however difficult to compare the sensitivity of different experiments. We therefore prefer to provide upper bounds on $\mu_A$ as a function of $\tau_A$ for fixed $m_A$.

In principle, this must be done independently for both $\mu_A = \text{BR}(B \to K A) \times \text{BR}(A \to \mu^+ \mu^-)$ and $\mu_A^\ast = \text{BR}(B \to K^\ast A) \times \text{BR}(A \to \mu^+ \mu^-)$, since the two B-meson branching ratios depend in different ways on the flavour-changing coefficients defined in eq.~(\ref{eq:hRhL}). Nevertheless, if the underlying interactions satisfy the hypothesis of minimal flavour violation~\cite{DAmbrosio:2002vsn}, the flavour-changing coefficients can always be written in the form of eq.~(\ref{eq:FCNC}) with an appropriate choice of $\Lambda$. As a result, both $\Gamma(B \to K^\ast \, A)$ and $\Gamma(B \to K \, A)$ are then proportional to $\log(\Lambda^2 / m_t^2)$, such that the ratio
\begin{align}
 \frac{\Gamma(B \rightarrow K^\ast \, A)}{\Gamma(B \rightarrow K \, A)} = & \frac{1}{(m_{B}^2 - m_{K}^2)^2} \frac{\left| A_0^{B^0}(m_A^2)\right|^2}{\left| f_0^{B^0}(m_A^2)\right|^2} \nonumber \\ & \times \frac{\lambda^{3/2}(m_{B}^2, m_{K^\ast}^2, m_A^2)}{\lambda^{1/2}(m_{B}^2, m_{K}^2, m_A^2)}\;, &
 \label{eq:ratio}
\end{align}
is independent of $\Lambda$ and does not contain any model-dependent quantities apart from $m_A$.

\begin{figure*}[t]
  \centering
  \includegraphics[width=0.4\textwidth]{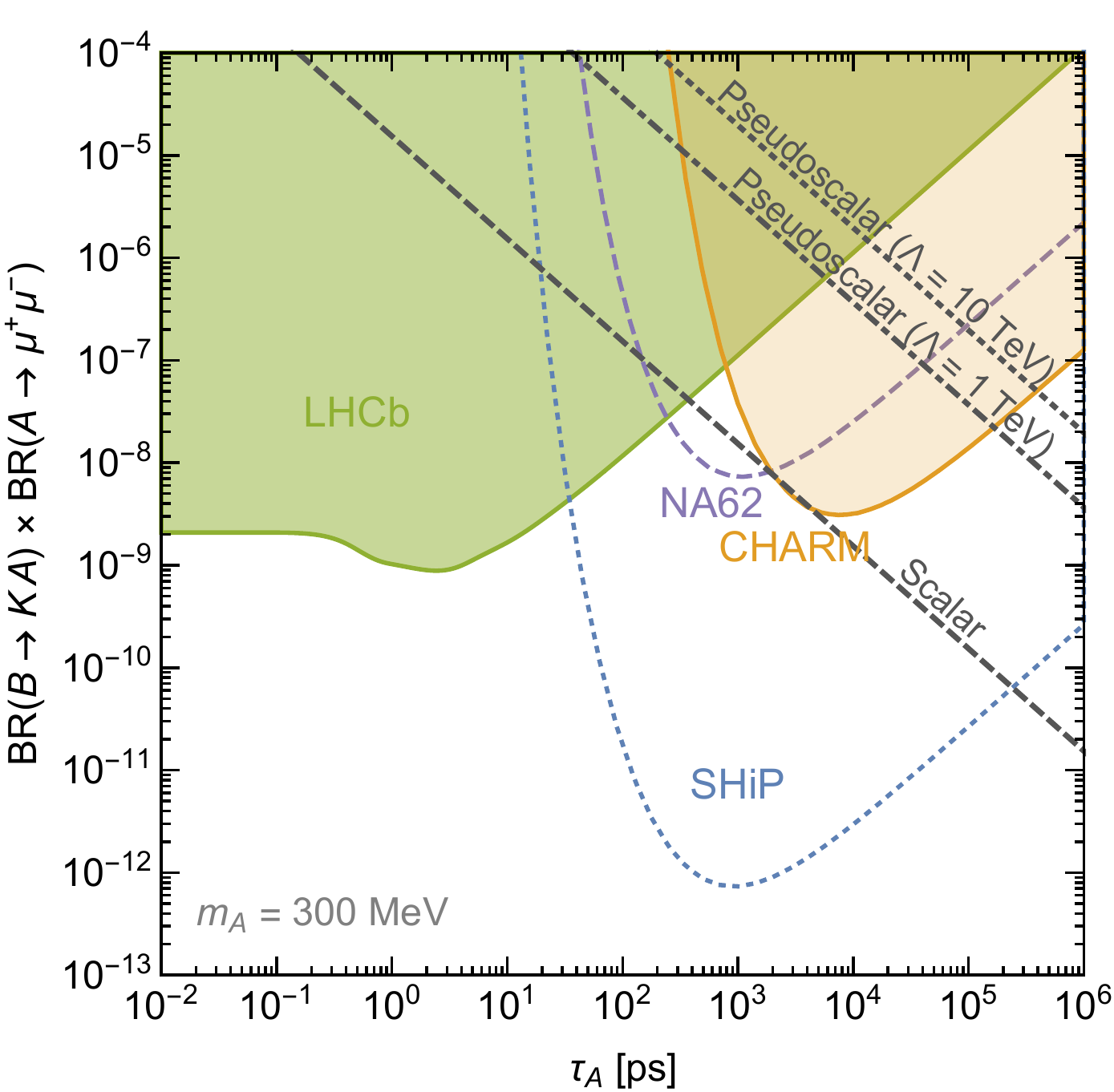}\qquad
  \includegraphics[width=0.4\textwidth]{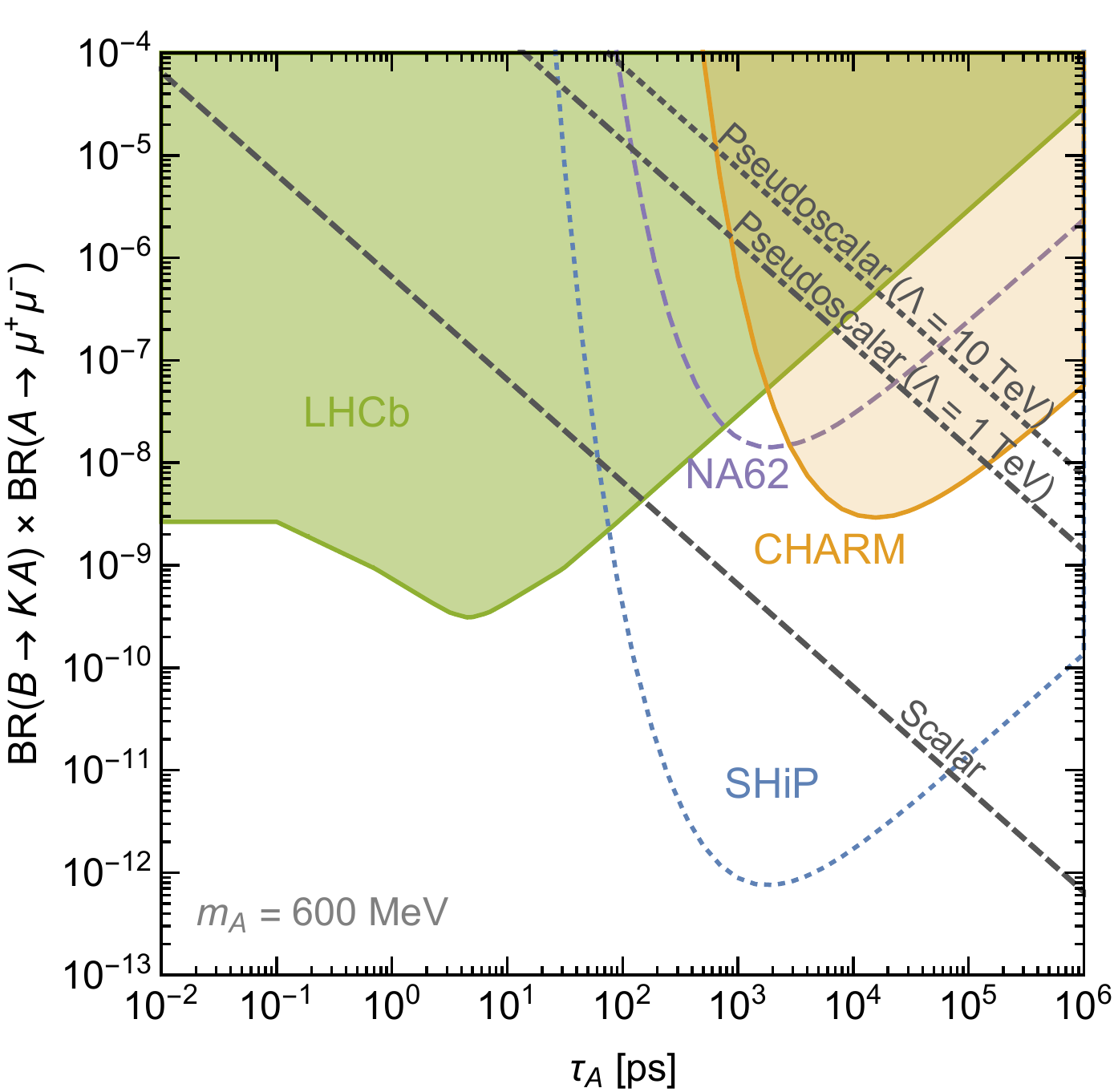}
  
  \includegraphics[width=0.4\textwidth]{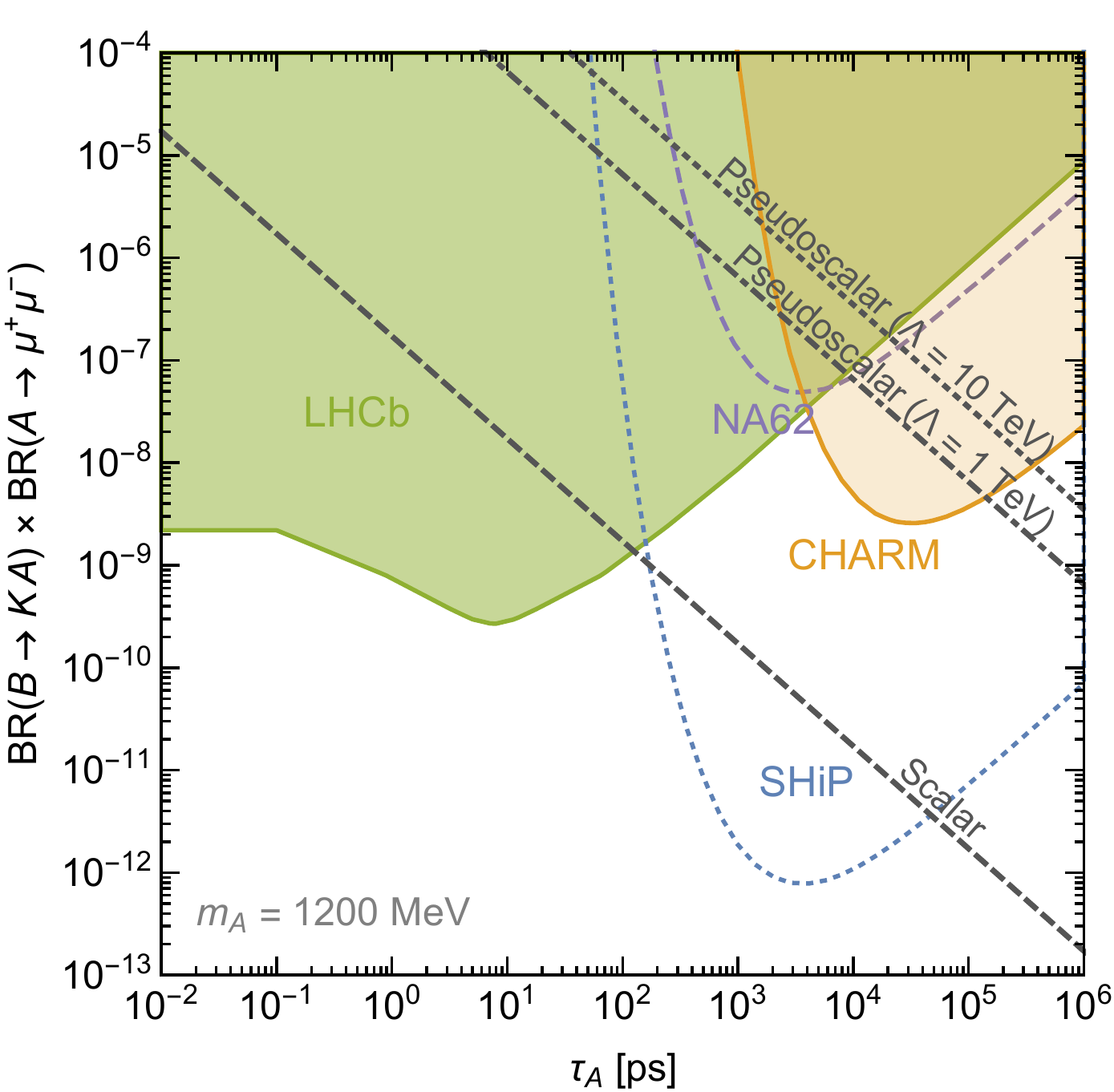}\qquad
  \includegraphics[width=0.4\textwidth]{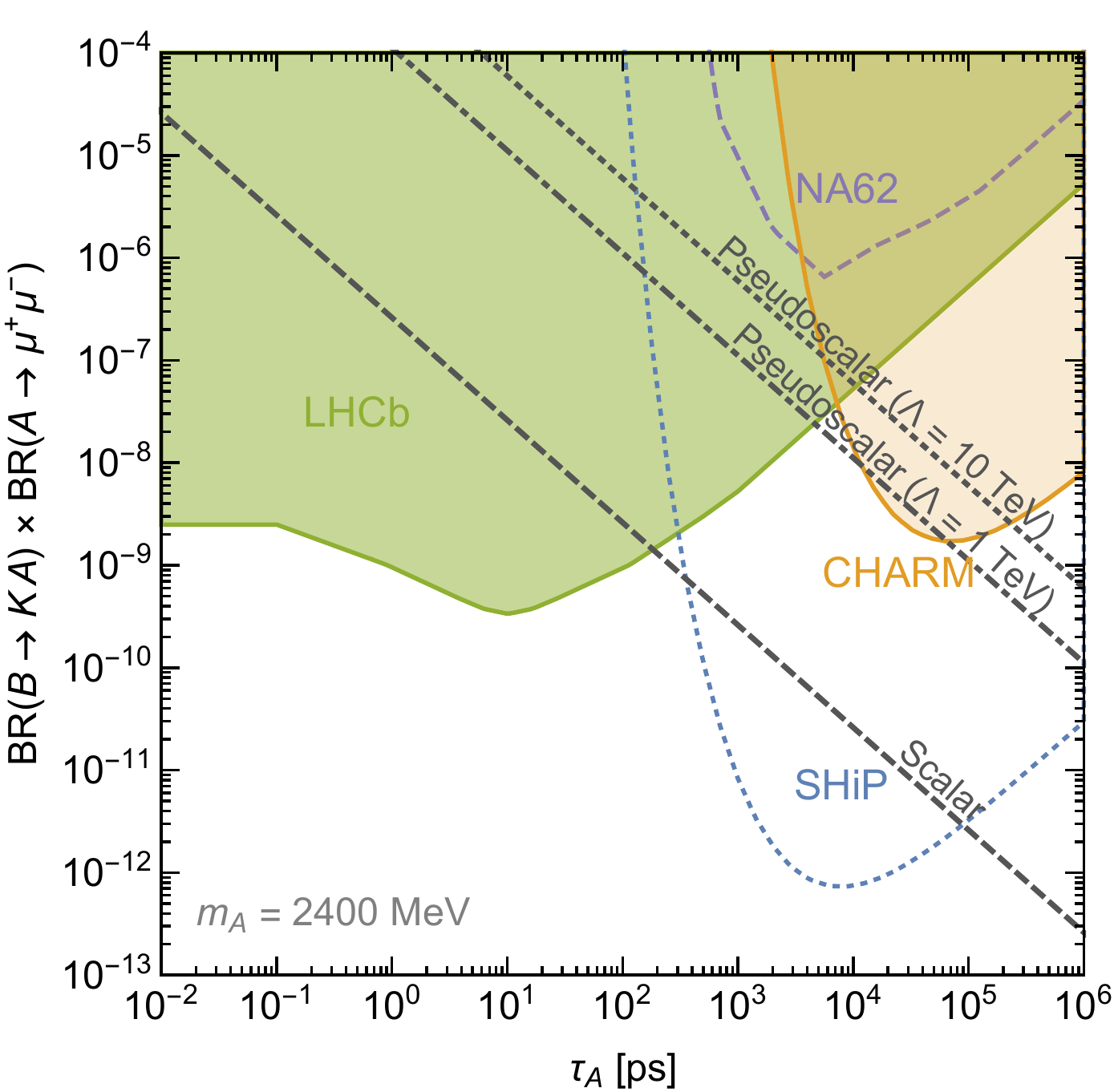}
  \vspace{-3mm}
  \caption{Model-independent constraints at 95\% confidence level on the production and decay of light scalars or pseudoscalars, expressed as bounds on 
  $\text{BR}(B \to KA)\times\text{BR}(A\to\mu^+\mu^-)$ as a function of the lifetime $\tau_A$ for different values of the boson mass $m_A$. 
  We also show for illustration the specific predictions obtained from the pseudoscalar model considered in section~\ref{sec:model} for different values of $\Lambda$, 
  as well as from the scalar model discussed in Ref.~\cite{Winkler:2018qyg}.}
  \label{fig:modelind}
\end{figure*}

For all models that satisfy minimal flavour violation, we can thus combine constraints on both $\mu_A$ and $\mu_A^\ast$ into a single plot. For the case of fixed-target experiments this is done by including both decay modes in the signal simulation, fixing the ratio between them according to eq.~(\ref{eq:ratio}). In the case of LHCb, on the other hand, we evaluate the constraints on $\mu_A$ and $\mu_A^\ast$ separately and show the stronger one.

We present our results in figure~\ref{fig:modelind}, where the four different panels correspond to different values of $m_A$. As in the bottom panel of figure~\ref{fig:results}, shaded regions correspond to existing exclusion bounds, while dashed and dotted regions indicate the sensitivity of planned experiments. We observe that the different experiments achieve their maximum sensitivity (i.e.\ the strongest bound on $\mu_A$) for different values of the lifetime $\tau_A$, reflecting the different geometries and different Lorentz-boost factors of the pseudoscalars. Indeed, the sensitivity is typically maximised if the pseudoscalar decay length corresponds roughly to the distance of the sensitive volume of the experiment from the production point. Hence, LHCb probes mostly short-lived pseudoscalars, whereas fixed-target experiments achieve the best sensitivity for longer lifetimes.

In particular, we find that the sensitivity of NA62 and CHARM peak at different lifetimes. This is because the B-meson energies, and hence the pseudoscalar Lorentz boosts, are comparable in both experiments but the distance of the sensitive volume is much shorter in NA62 than in CHARM. We see that this different geometry enables NA62  to probe low-mass pseudoscalars with lifetimes $100\,\mathrm{ps} \leq \mu_A \leq 1000\,\mathrm{ps}$, a region for which neither LHCb nor CHARM are very sensitive. This finding is in stark contrast with the conclusions of figure~\ref{fig:results}, where we found NA62 unable to provide relevant constraints. In other words, the model-independent analysis reveals the unique potential of NA62 to probe certain regions of parameter space.

Figure~\ref{fig:modelind} also contains a number of lines corresponding to the predicted relation between $\mu_A$ and $\tau_A$ in specific models. The line labelled ``Pseudoscalar ($\Lambda = 1\,\mathrm{TeV}$)'' corresponds to the model discussed in Secs.~\ref{sec:model} and~\ref{sec:experiments} (see in particular figure~\ref{fig:results}). Consistent with our observations there we find that with this assumption NA62 does not probe any parameter combinations that are not already excluded by the combination of LHCb and CHARM.
This conclusion also does not change when changing the value of $\Lambda$ assumed in the calculation of B-meson decays by moderate amounts. Increasing $\Lambda$ simply shifts the line towards larger values of $\mu_A$, such that the constraints from LHCb and CHARM become even stronger (and vice versa). Adding an invisible decay channel would shift the line to the bottom-left (as both the lifetime and the leptonic branching ratios are decreased), such that experimental constraints can be evaded.

A key advantage of the model-independent approach is that we did not need to assume at any point that the light boson is a pseudoscalar (i.e.\ CP-odd). Indeed, the constraints that we show apply equally to light scalars that satisfy minimal flavour violation (see e.g.\ Refs.~\cite{Clarke:2013aya,Dev:2017dui,Winkler:2018qyg}). For illustration, we also indicate in figure~\ref{fig:modelind} the model-specific predictions for a light scalar that mixes with the Higgs boson (studied most recently in Ref.~\cite{Winkler:2018qyg}). This model has the advantage that the loop-induced flavour-changing coefficients are finite and hence independent of the new-physics scale $\Lambda$. The calculation of the scalar lifetime and branching ratios, on the other hand, is much more challenging, as hadronic decay modes cannot be neglected. The combination of these two effects shifts the line for the scalar case to the bottom and to the left. Indeed, we find that NA62 may possess a unique sensitivity to this model for scalar lifetimes in the range $100\text{--}1000\,\mathrm{ps}$.

\section{Conclusions}
\label{sec:conclusions}

Light pseudoscalars arise naturally in many extensions of the Standard Model and may be an important ingredient of phenomenologically viable dark matter models. Important constraints on these models arise from searches for long-lived particles in fixed-target experiments and from searches for rare B-meson decays at LHCb. In this work we have improved the analysis of existing constraints from these experiments and proposed a new method for comparing the sensitivities of future searches. 

For the first part of the paper we have adopted an effective low-energy description of the interactions of pseudoscalars with quarks and charged leptons. Loop-induced flavour-changing processes then induce a sensitivity to new physics at higher scales, parametrised by the unknown scale $\Lambda$, which enters in the production rate of light pseudoscalars via rare B-meson decays. Evaluating experimental constraints for fixed values of $\Lambda$, we find that searches for displaced vertices at LHCb and an improved calculation of the constraints from CHARM in combination rule out much of the relevant parameter space.

Nevertheless, the unknown relation between the rates of production and decay of light pseudoscalars (together with additional theoretical uncertainties in the signal calculation) makes it desirable to adopt a more model-independent approach. To this end we propose to express experimental results and sensitivities as bounds on products of branching ratios like $\mu_A = \text{BR}(B \to K\,A) \times \text{BR}(A \to \mu^+\mu^-)$ as a function of the pseudoscalar lifetime $\tau_A$ and mass $m_A$. This form of presentation has the additional advantage that it makes the role of different experimental geometries explicit, because the sensitivity is typically maximised for pseudoscalar decay lengths comparable to the size of the experiment.

Adopting this model-independent approach, we demonstrate that NA62 has potentially world-leading sensitivity for pseudoscalar lifetimes in the range of $100\text{--}1000\,\mathrm{ps}$. Further substantial progress can be expected from dedicated facilities to search for long-lived particles like SHiP. Our approach can easily be generalised to include other pseudoscalar decay modes, such as $A \to \gamma \gamma$ or $A \to 3\pi$, which yield complementary information. In summary, while there exist already strong constraints on light pseudoscalars, substantial terrain remains unexplored and provides room for further exploration and potential discoveries.

\section*{Acknowledgements}

We thank Herbi Dreiner, Florian Domingo, Pedro Schwaller and Peter Skands for discussions, Maxim Pospelov for the encouragement to revise the CHARM limits and Martin W.~Winkler for helpful comments on the manuscript and for providing digitised results from Ref.~\cite{Winkler:2018qyg}. BD acknowledges support through ERC-STG 802836 (AxScale). FE and FK are funded by the DFG Emmy Noether Grant No.\ KA 4662/1-1.


\begin{thebibliography}{70}%
\makeatletter
\providecommand \@ifxundefined [1]{%
 \@ifx{#1\undefined}
}%
\providecommand \@ifnum [1]{%
 \ifnum #1\expandafter \@firstoftwo
 \else \expandafter \@secondoftwo
 \fi
}%
\providecommand \@ifx [1]{%
 \ifx #1\expandafter \@firstoftwo
 \else \expandafter \@secondoftwo
 \fi
}%
\providecommand \natexlab [1]{#1}%
\providecommand \enquote  [1]{``#1''}%
\providecommand \bibnamefont  [1]{#1}%
\providecommand \bibfnamefont [1]{#1}%
\providecommand \citenamefont [1]{#1}%
\providecommand \href@noop [0]{\@secondoftwo}%
\providecommand \href [0]{\begingroup \@sanitize@url \@href}%
\providecommand \@href[1]{\@@startlink{#1}\@@href}%
\providecommand \@@href[1]{\endgroup#1\@@endlink}%
\providecommand \@sanitize@url [0]{\catcode `\\12\catcode `\$12\catcode
  `\&12\catcode `\#12\catcode `\^12\catcode `\_12\catcode `\%12\relax}%
\providecommand \@@startlink[1]{}%
\providecommand \@@endlink[0]{}%
\providecommand \url  [0]{\begingroup\@sanitize@url \@url }%
\providecommand \@url [1]{\endgroup\@href {#1}{\urlprefix }}%
\providecommand \urlprefix  [0]{URL }%
\providecommand \Eprint [0]{\href }%
\providecommand \doibase [0]{http://dx.doi.org/}%
\providecommand \selectlanguage [0]{\@gobble}%
\providecommand \bibinfo  [0]{\@secondoftwo}%
\providecommand \bibfield  [0]{\@secondoftwo}%
\providecommand \translation [1]{[#1]}%
\providecommand \BibitemOpen [0]{}%
\providecommand \bibitemStop [0]{}%
\providecommand \bibitemNoStop [0]{.\EOS\space}%
\providecommand \EOS [0]{\spacefactor3000\relax}%
\providecommand \BibitemShut  [1]{\csname bibitem#1\endcsname}%
\let\auto@bib@innerbib\@empty
\bibitem [{\citenamefont {Nomura}\ and\ \citenamefont
  {Thaler}(2009)}]{Nomura:2008ru}%
  \BibitemOpen
  \bibfield  {author} {\bibinfo {author} {\bibfnamefont {Y.}~\bibnamefont
  {Nomura}}\ and\ \bibinfo {author} {\bibfnamefont {J.}~\bibnamefont
  {Thaler}},\ }\href {\doibase 10.1103/PhysRevD.79.075008} {\bibfield
  {journal} {\bibinfo  {journal} {Phys. Rev.}\ }\textbf {\bibinfo {volume}
  {D79}},\ \bibinfo {pages} {075008} (\bibinfo {year} {2009})},\ \Eprint
  {http://arxiv.org/abs/0810.5397} {arXiv:0810.5397 [hep-ph]} \BibitemShut
  {NoStop}%
\bibitem [{\citenamefont {Batell}\ \emph {et~al.}(2009)\citenamefont {Batell},
  \citenamefont {Pospelov},\ and\ \citenamefont {Ritz}}]{Batell:2009di}%
  \BibitemOpen
  \bibfield  {author} {\bibinfo {author} {\bibfnamefont {B.}~\bibnamefont
  {Batell}}, \bibinfo {author} {\bibfnamefont {M.}~\bibnamefont {Pospelov}}, \
  and\ \bibinfo {author} {\bibfnamefont {A.}~\bibnamefont {Ritz}},\ }\href
  {\doibase 10.1103/PhysRevD.80.095024} {\bibfield  {journal} {\bibinfo
  {journal} {Phys. Rev.}\ }\textbf {\bibinfo {volume} {D80}},\ \bibinfo {pages}
  {095024} (\bibinfo {year} {2009})},\ \Eprint {http://arxiv.org/abs/0906.5614}
  {arXiv:0906.5614 [hep-ph]} \BibitemShut {NoStop}%
\bibitem [{\citenamefont {Freytsis}\ \emph {et~al.}(2010)\citenamefont
  {Freytsis}, \citenamefont {Ligeti},\ and\ \citenamefont
  {Thaler}}]{Freytsis:2009ct}%
  \BibitemOpen
  \bibfield  {author} {\bibinfo {author} {\bibfnamefont {M.}~\bibnamefont
  {Freytsis}}, \bibinfo {author} {\bibfnamefont {Z.}~\bibnamefont {Ligeti}}, \
  and\ \bibinfo {author} {\bibfnamefont {J.}~\bibnamefont {Thaler}},\ }\href
  {\doibase 10.1103/PhysRevD.81.034001} {\bibfield  {journal} {\bibinfo
  {journal} {Phys. Rev.}\ }\textbf {\bibinfo {volume} {D81}},\ \bibinfo {pages}
  {034001} (\bibinfo {year} {2010})},\ \Eprint {http://arxiv.org/abs/0911.5355}
  {arXiv:0911.5355 [hep-ph]} \BibitemShut {NoStop}%
\bibitem [{\citenamefont {Hochberg}\ \emph {et~al.}(2018)\citenamefont
  {Hochberg}, \citenamefont {Kuflik}, \citenamefont {Mcgehee}, \citenamefont
  {Murayama},\ and\ \citenamefont {Schutz}}]{Hochberg:2018rjs}%
  \BibitemOpen
  \bibfield  {author} {\bibinfo {author} {\bibfnamefont {Y.}~\bibnamefont
  {Hochberg}}, \bibinfo {author} {\bibfnamefont {E.}~\bibnamefont {Kuflik}},
  \bibinfo {author} {\bibfnamefont {R.}~\bibnamefont {Mcgehee}}, \bibinfo
  {author} {\bibfnamefont {H.}~\bibnamefont {Murayama}}, \ and\ \bibinfo
  {author} {\bibfnamefont {K.}~\bibnamefont {Schutz}},\ }\href@noop {} {\
  (\bibinfo {year} {2018})},\ \Eprint {http://arxiv.org/abs/1806.10139}
  {arXiv:1806.10139 [hep-ph]} \BibitemShut {NoStop}%
\bibitem [{\citenamefont {Mimasu}\ and\ \citenamefont
  {Sanz}(2015)}]{Mimasu:2014nea}%
  \BibitemOpen
  \bibfield  {author} {\bibinfo {author} {\bibfnamefont {K.}~\bibnamefont
  {Mimasu}}\ and\ \bibinfo {author} {\bibfnamefont {V.}~\bibnamefont {Sanz}},\
  }\href {\doibase 10.1007/JHEP06(2015)173} {\bibfield  {journal} {\bibinfo
  {journal} {JHEP}\ }\textbf {\bibinfo {volume} {06}},\ \bibinfo {pages} {173}
  (\bibinfo {year} {2015})},\ \Eprint {http://arxiv.org/abs/1409.4792}
  {arXiv:1409.4792 [hep-ph]} \BibitemShut {NoStop}%
\bibitem [{\citenamefont {Jaeckel}\ and\ \citenamefont
  {Spannowsky}(2016)}]{Jaeckel:2015jla}%
  \BibitemOpen
  \bibfield  {author} {\bibinfo {author} {\bibfnamefont {J.}~\bibnamefont
  {Jaeckel}}\ and\ \bibinfo {author} {\bibfnamefont {M.}~\bibnamefont
  {Spannowsky}},\ }\href {\doibase 10.1016/j.physletb.2015.12.037} {\bibfield
  {journal} {\bibinfo  {journal} {Phys. Lett.}\ }\textbf {\bibinfo {volume}
  {B753}},\ \bibinfo {pages} {482} (\bibinfo {year} {2016})},\ \Eprint
  {http://arxiv.org/abs/1509.00476} {arXiv:1509.00476 [hep-ph]} \BibitemShut
  {NoStop}%
\bibitem [{\citenamefont {Brivio}\ \emph {et~al.}(2017)\citenamefont {Brivio},
  \citenamefont {Gavela}, \citenamefont {Merlo}, \citenamefont {Mimasu},
  \citenamefont {No}, \citenamefont {del Rey},\ and\ \citenamefont
  {Sanz}}]{Brivio:2017ije}%
  \BibitemOpen
  \bibfield  {author} {\bibinfo {author} {\bibfnamefont {I.}~\bibnamefont
  {Brivio}}, \bibinfo {author} {\bibfnamefont {M.~B.}\ \bibnamefont {Gavela}},
  \bibinfo {author} {\bibfnamefont {L.}~\bibnamefont {Merlo}}, \bibinfo
  {author} {\bibfnamefont {K.}~\bibnamefont {Mimasu}}, \bibinfo {author}
  {\bibfnamefont {J.~M.}\ \bibnamefont {No}}, \bibinfo {author} {\bibfnamefont
  {R.}~\bibnamefont {del Rey}}, \ and\ \bibinfo {author} {\bibfnamefont
  {V.}~\bibnamefont {Sanz}},\ }\href {\doibase 10.1140/epjc/s10052-017-5111-3}
  {\bibfield  {journal} {\bibinfo  {journal} {Eur. Phys. J.}\ }\textbf
  {\bibinfo {volume} {C77}},\ \bibinfo {pages} {572} (\bibinfo {year}
  {2017})},\ \Eprint {http://arxiv.org/abs/1701.05379} {arXiv:1701.05379
  [hep-ph]} \BibitemShut {NoStop}%
\bibitem [{\citenamefont {Bauer}\ \emph
  {et~al.}(2017{\natexlab{a}})\citenamefont {Bauer}, \citenamefont {Neubert},\
  and\ \citenamefont {Thamm}}]{Bauer:2017ris}%
  \BibitemOpen
  \bibfield  {author} {\bibinfo {author} {\bibfnamefont {M.}~\bibnamefont
  {Bauer}}, \bibinfo {author} {\bibfnamefont {M.}~\bibnamefont {Neubert}}, \
  and\ \bibinfo {author} {\bibfnamefont {A.}~\bibnamefont {Thamm}},\ }\href
  {\doibase 10.1007/JHEP12(2017)044} {\bibfield  {journal} {\bibinfo  {journal}
  {JHEP}\ }\textbf {\bibinfo {volume} {12}},\ \bibinfo {pages} {044} (\bibinfo
  {year} {2017}{\natexlab{a}})},\ \Eprint {http://arxiv.org/abs/1708.00443}
  {arXiv:1708.00443 [hep-ph]} \BibitemShut {NoStop}%
\bibitem [{\citenamefont {Haisch}\ \emph {et~al.}(2018)\citenamefont {Haisch},
  \citenamefont {Kamenik}, \citenamefont {Malinauskas},\ and\ \citenamefont
  {Spira}}]{Haisch:2018kqx}%
  \BibitemOpen
  \bibfield  {author} {\bibinfo {author} {\bibfnamefont {U.}~\bibnamefont
  {Haisch}}, \bibinfo {author} {\bibfnamefont {J.~F.}\ \bibnamefont {Kamenik}},
  \bibinfo {author} {\bibfnamefont {A.}~\bibnamefont {Malinauskas}}, \ and\
  \bibinfo {author} {\bibfnamefont {M.}~\bibnamefont {Spira}},\ }\href
  {\doibase 10.1007/JHEP03(2018)178} {\bibfield  {journal} {\bibinfo  {journal}
  {JHEP}\ }\textbf {\bibinfo {volume} {03}},\ \bibinfo {pages} {178} (\bibinfo
  {year} {2018})},\ \Eprint {http://arxiv.org/abs/1802.02156} {arXiv:1802.02156
  [hep-ph]} \BibitemShut {NoStop}%
\bibitem [{\citenamefont {Andreas}\ \emph {et~al.}(2010)\citenamefont
  {Andreas}, \citenamefont {Lebedev}, \citenamefont {Ramos-Sanchez},\ and\
  \citenamefont {Ringwald}}]{Andreas:2010ms}%
  \BibitemOpen
  \bibfield  {author} {\bibinfo {author} {\bibfnamefont {S.}~\bibnamefont
  {Andreas}}, \bibinfo {author} {\bibfnamefont {O.}~\bibnamefont {Lebedev}},
  \bibinfo {author} {\bibfnamefont {S.}~\bibnamefont {Ramos-Sanchez}}, \ and\
  \bibinfo {author} {\bibfnamefont {A.}~\bibnamefont {Ringwald}},\ }\href
  {\doibase 10.1007/JHEP08(2010)003} {\bibfield  {journal} {\bibinfo  {journal}
  {JHEP}\ }\textbf {\bibinfo {volume} {1008}},\ \bibinfo {pages} {003}
  (\bibinfo {year} {2010})},\ \Eprint {http://arxiv.org/abs/1005.3978}
  {arXiv:1005.3978 [hep-ph]} \BibitemShut {NoStop}%
\bibitem [{\citenamefont {Hewett}\ \emph {et~al.}(2012)\citenamefont {Hewett}
  \emph {et~al.}}]{Hewett:2012ns}%
  \BibitemOpen
  \bibfield  {author} {\bibinfo {author} {\bibfnamefont {J.~L.}\ \bibnamefont
  {Hewett}} \emph {et~al.},\ }\href@noop {} {\  (\bibinfo {year} {2012})},\
  \Eprint {http://arxiv.org/abs/1205.2671} {arXiv:1205.2671 [hep-ex]}
  \BibitemShut {NoStop}%
\bibitem [{\citenamefont {Essig}\ \emph {et~al.}(2013)\citenamefont {Essig}
  \emph {et~al.}}]{Essig:2013lka}%
  \BibitemOpen
  \bibfield  {author} {\bibinfo {author} {\bibfnamefont {R.}~\bibnamefont
  {Essig}} \emph {et~al.},\ }\href@noop {} {\  (\bibinfo {year} {2013})},\
  \Eprint {http://arxiv.org/abs/1311.0029} {arXiv:1311.0029 [hep-ph]}
  \BibitemShut {NoStop}%
\bibitem [{\citenamefont {Flacke}\ \emph {et~al.}(2017)\citenamefont {Flacke},
  \citenamefont {Frugiuele}, \citenamefont {Fuchs}, \citenamefont {Gupta},\
  and\ \citenamefont {Perez}}]{Flacke:2016szy}%
  \BibitemOpen
  \bibfield  {author} {\bibinfo {author} {\bibfnamefont {T.}~\bibnamefont
  {Flacke}}, \bibinfo {author} {\bibfnamefont {C.}~\bibnamefont {Frugiuele}},
  \bibinfo {author} {\bibfnamefont {E.}~\bibnamefont {Fuchs}}, \bibinfo
  {author} {\bibfnamefont {R.~S.}\ \bibnamefont {Gupta}}, \ and\ \bibinfo
  {author} {\bibfnamefont {G.}~\bibnamefont {Perez}},\ }\href {\doibase
  10.1007/JHEP06(2017)050} {\bibfield  {journal} {\bibinfo  {journal} {JHEP}\
  }\textbf {\bibinfo {volume} {06}},\ \bibinfo {pages} {050} (\bibinfo {year}
  {2017})},\ \Eprint {http://arxiv.org/abs/1610.02025} {arXiv:1610.02025
  [hep-ph]} \BibitemShut {NoStop}%
\bibitem [{\citenamefont {Freytsis}\ and\ \citenamefont
  {Ligeti}(2011)}]{Freytsis:2010ne}%
  \BibitemOpen
  \bibfield  {author} {\bibinfo {author} {\bibfnamefont {M.}~\bibnamefont
  {Freytsis}}\ and\ \bibinfo {author} {\bibfnamefont {Z.}~\bibnamefont
  {Ligeti}},\ }\href {\doibase 10.1103/PhysRevD.83.115009} {\bibfield
  {journal} {\bibinfo  {journal} {Phys.Rev.}\ }\textbf {\bibinfo {volume}
  {D83}},\ \bibinfo {pages} {115009} (\bibinfo {year} {2011})},\ \Eprint
  {http://arxiv.org/abs/1012.5317} {arXiv:1012.5317 [hep-ph]} \BibitemShut
  {NoStop}%
\bibitem [{\citenamefont {Dienes}\ \emph {et~al.}(2014)\citenamefont {Dienes},
  \citenamefont {Kumar}, \citenamefont {Thomas},\ and\ \citenamefont
  {Yaylali}}]{Dienes:2013xya}%
  \BibitemOpen
  \bibfield  {author} {\bibinfo {author} {\bibfnamefont {K.~R.}\ \bibnamefont
  {Dienes}}, \bibinfo {author} {\bibfnamefont {J.}~\bibnamefont {Kumar}},
  \bibinfo {author} {\bibfnamefont {B.}~\bibnamefont {Thomas}}, \ and\ \bibinfo
  {author} {\bibfnamefont {D.}~\bibnamefont {Yaylali}},\ }\href {\doibase
  10.1103/PhysRevD.90.015012} {\bibfield  {journal} {\bibinfo  {journal}
  {Phys.Rev.}\ }\textbf {\bibinfo {volume} {D90}},\ \bibinfo {pages} {015012}
  (\bibinfo {year} {2014})},\ \Eprint {http://arxiv.org/abs/1312.7772}
  {arXiv:1312.7772 [hep-ph]} \BibitemShut {NoStop}%
\bibitem [{\citenamefont {Arina}\ \emph {et~al.}(2014)\citenamefont {Arina},
  \citenamefont {Del~Nobile},\ and\ \citenamefont {Panci}}]{Arina:2014yna}%
  \BibitemOpen
  \bibfield  {author} {\bibinfo {author} {\bibfnamefont {C.}~\bibnamefont
  {Arina}}, \bibinfo {author} {\bibfnamefont {E.}~\bibnamefont {Del~Nobile}}, \
  and\ \bibinfo {author} {\bibfnamefont {P.}~\bibnamefont {Panci}},\
  }\href@noop {} {\  (\bibinfo {year} {2014})},\ \Eprint
  {http://arxiv.org/abs/1406.5542} {arXiv:1406.5542 [hep-ph]} \BibitemShut
  {NoStop}%
\bibitem [{\citenamefont {Arcadi}\ \emph {et~al.}(2018)\citenamefont {Arcadi},
  \citenamefont {Lindner}, \citenamefont {Queiroz}, \citenamefont
  {Rodejohann},\ and\ \citenamefont {Vogl}}]{Arcadi:2017wqi}%
  \BibitemOpen
  \bibfield  {author} {\bibinfo {author} {\bibfnamefont {G.}~\bibnamefont
  {Arcadi}}, \bibinfo {author} {\bibfnamefont {M.}~\bibnamefont {Lindner}},
  \bibinfo {author} {\bibfnamefont {F.~S.}\ \bibnamefont {Queiroz}}, \bibinfo
  {author} {\bibfnamefont {W.}~\bibnamefont {Rodejohann}}, \ and\ \bibinfo
  {author} {\bibfnamefont {S.}~\bibnamefont {Vogl}},\ }\href {\doibase
  10.1088/1475-7516/2018/03/042} {\bibfield  {journal} {\bibinfo  {journal}
  {JCAP}\ }\textbf {\bibinfo {volume} {1803}},\ \bibinfo {pages} {042}
  (\bibinfo {year} {2018})},\ \Eprint {http://arxiv.org/abs/1711.02110}
  {arXiv:1711.02110 [hep-ph]} \BibitemShut {NoStop}%
\bibitem [{\citenamefont {Bell}\ \emph {et~al.}(2018)\citenamefont {Bell},
  \citenamefont {Busoni},\ and\ \citenamefont {Sanderson}}]{Sanderson:2018lmj}%
  \BibitemOpen
  \bibfield  {author} {\bibinfo {author} {\bibfnamefont {N.~F.}\ \bibnamefont
  {Bell}}, \bibinfo {author} {\bibfnamefont {G.}~\bibnamefont {Busoni}}, \ and\
  \bibinfo {author} {\bibfnamefont {I.~W.}\ \bibnamefont {Sanderson}},\ }\href
  {\doibase 10.1088/1475-7516/2018/08/017} {\bibfield  {journal} {\bibinfo
  {journal} {JCAP}\ }\textbf {\bibinfo {volume} {1808}},\ \bibinfo {pages}
  {017} (\bibinfo {year} {2018})},\ \Eprint {http://arxiv.org/abs/1803.01574}
  {arXiv:1803.01574 [hep-ph]} \BibitemShut {NoStop}%
\bibitem [{\citenamefont {Abe}\ \emph {et~al.}(2018)\citenamefont {Abe},
  \citenamefont {Fujiwara},\ and\ \citenamefont {Hisano}}]{Abe:2018emu}%
  \BibitemOpen
  \bibfield  {author} {\bibinfo {author} {\bibfnamefont {T.}~\bibnamefont
  {Abe}}, \bibinfo {author} {\bibfnamefont {M.}~\bibnamefont {Fujiwara}}, \
  and\ \bibinfo {author} {\bibfnamefont {J.}~\bibnamefont {Hisano}},\
  }\href@noop {} {\  (\bibinfo {year} {2018})},\ \Eprint
  {http://arxiv.org/abs/1810.01039} {arXiv:1810.01039 [hep-ph]} \BibitemShut
  {NoStop}%
\bibitem [{\citenamefont {Boehm}\ \emph {et~al.}(2014)\citenamefont {Boehm},
  \citenamefont {Dolan}, \citenamefont {McCabe}, \citenamefont {Spannowsky},\
  and\ \citenamefont {Wallace}}]{Boehm:2014hva}%
  \BibitemOpen
  \bibfield  {author} {\bibinfo {author} {\bibfnamefont {C.}~\bibnamefont
  {Boehm}}, \bibinfo {author} {\bibfnamefont {M.~J.}\ \bibnamefont {Dolan}},
  \bibinfo {author} {\bibfnamefont {C.}~\bibnamefont {McCabe}}, \bibinfo
  {author} {\bibfnamefont {M.}~\bibnamefont {Spannowsky}}, \ and\ \bibinfo
  {author} {\bibfnamefont {C.~J.}\ \bibnamefont {Wallace}},\ }\href {\doibase
  10.1088/1475-7516/2014/05/009} {\bibfield  {journal} {\bibinfo  {journal}
  {JCAP}\ }\textbf {\bibinfo {volume} {1405}},\ \bibinfo {pages} {009}
  (\bibinfo {year} {2014})},\ \Eprint {http://arxiv.org/abs/1401.6458}
  {arXiv:1401.6458 [hep-ph]} \BibitemShut {NoStop}%
\bibitem [{\citenamefont {Ipek}\ \emph {et~al.}(2014)\citenamefont {Ipek},
  \citenamefont {McKeen},\ and\ \citenamefont {Nelson}}]{Ipek:2014gua}%
  \BibitemOpen
  \bibfield  {author} {\bibinfo {author} {\bibfnamefont {S.}~\bibnamefont
  {Ipek}}, \bibinfo {author} {\bibfnamefont {D.}~\bibnamefont {McKeen}}, \ and\
  \bibinfo {author} {\bibfnamefont {A.~E.}\ \bibnamefont {Nelson}},\ }\href
  {\doibase 10.1103/PhysRevD.90.055021} {\bibfield  {journal} {\bibinfo
  {journal} {Phys.Rev.}\ }\textbf {\bibinfo {volume} {D90}},\ \bibinfo {pages}
  {055021} (\bibinfo {year} {2014})},\ \Eprint {http://arxiv.org/abs/1404.3716}
  {arXiv:1404.3716 [hep-ph]} \BibitemShut {NoStop}%
\bibitem [{\citenamefont {Berlin}\ \emph {et~al.}(2015)\citenamefont {Berlin},
  \citenamefont {Gori}, \citenamefont {Lin},\ and\ \citenamefont
  {Wang}}]{Berlin:2015wwa}%
  \BibitemOpen
  \bibfield  {author} {\bibinfo {author} {\bibfnamefont {A.}~\bibnamefont
  {Berlin}}, \bibinfo {author} {\bibfnamefont {S.}~\bibnamefont {Gori}},
  \bibinfo {author} {\bibfnamefont {T.}~\bibnamefont {Lin}}, \ and\ \bibinfo
  {author} {\bibfnamefont {L.-T.}\ \bibnamefont {Wang}},\ }\href {\doibase
  10.1103/PhysRevD.92.015005} {\bibfield  {journal} {\bibinfo  {journal} {Phys.
  Rev.}\ }\textbf {\bibinfo {volume} {D92}},\ \bibinfo {pages} {015005}
  (\bibinfo {year} {2015})},\ \Eprint {http://arxiv.org/abs/1502.06000}
  {arXiv:1502.06000 [hep-ph]} \BibitemShut {NoStop}%
\bibitem [{\citenamefont {Tunney}\ \emph {et~al.}(2017)\citenamefont {Tunney},
  \citenamefont {No},\ and\ \citenamefont {Fairbairn}}]{Tunney:2017yfp}%
  \BibitemOpen
  \bibfield  {author} {\bibinfo {author} {\bibfnamefont {P.}~\bibnamefont
  {Tunney}}, \bibinfo {author} {\bibfnamefont {J.~M.}\ \bibnamefont {No}}, \
  and\ \bibinfo {author} {\bibfnamefont {M.}~\bibnamefont {Fairbairn}},\ }\href
  {\doibase 10.1103/PhysRevD.96.095020} {\bibfield  {journal} {\bibinfo
  {journal} {Phys. Rev.}\ }\textbf {\bibinfo {volume} {D96}},\ \bibinfo {pages}
  {095020} (\bibinfo {year} {2017})},\ \Eprint
  {http://arxiv.org/abs/1705.09670} {arXiv:1705.09670 [hep-ph]} \BibitemShut
  {NoStop}%
\bibitem [{\citenamefont {No}(2016)}]{No:2015xqa}%
  \BibitemOpen
  \bibfield  {author} {\bibinfo {author} {\bibfnamefont {J.~M.}\ \bibnamefont
  {No}},\ }\href {\doibase 10.1103/PhysRevD.93.031701} {\bibfield  {journal}
  {\bibinfo  {journal} {Phys. Rev.}\ }\textbf {\bibinfo {volume} {D93}},\
  \bibinfo {pages} {031701} (\bibinfo {year} {2016})},\ \Eprint
  {http://arxiv.org/abs/1509.01110} {arXiv:1509.01110 [hep-ph]} \BibitemShut
  {NoStop}%
\bibitem [{\citenamefont {Goncalves}\ \emph {et~al.}(2017)\citenamefont
  {Goncalves}, \citenamefont {Machado},\ and\ \citenamefont
  {No}}]{Goncalves:2016iyg}%
  \BibitemOpen
  \bibfield  {author} {\bibinfo {author} {\bibfnamefont {D.}~\bibnamefont
  {Goncalves}}, \bibinfo {author} {\bibfnamefont {P.~A.~N.}\ \bibnamefont
  {Machado}}, \ and\ \bibinfo {author} {\bibfnamefont {J.~M.}\ \bibnamefont
  {No}},\ }\href {\doibase 10.1103/PhysRevD.95.055027} {\bibfield  {journal}
  {\bibinfo  {journal} {Phys. Rev.}\ }\textbf {\bibinfo {volume} {D95}},\
  \bibinfo {pages} {055027} (\bibinfo {year} {2017})},\ \Eprint
  {http://arxiv.org/abs/1611.04593} {arXiv:1611.04593 [hep-ph]} \BibitemShut
  {NoStop}%
\bibitem [{\citenamefont {Bauer}\ \emph
  {et~al.}(2017{\natexlab{b}})\citenamefont {Bauer}, \citenamefont {Haisch},\
  and\ \citenamefont {Kahlhoefer}}]{Bauer:2017ota}%
  \BibitemOpen
  \bibfield  {author} {\bibinfo {author} {\bibfnamefont {M.}~\bibnamefont
  {Bauer}}, \bibinfo {author} {\bibfnamefont {U.}~\bibnamefont {Haisch}}, \
  and\ \bibinfo {author} {\bibfnamefont {F.}~\bibnamefont {Kahlhoefer}},\
  }\href {\doibase 10.1007/JHEP05(2017)138} {\bibfield  {journal} {\bibinfo
  {journal} {JHEP}\ }\textbf {\bibinfo {volume} {05}},\ \bibinfo {pages} {138}
  (\bibinfo {year} {2017}{\natexlab{b}})},\ \Eprint
  {http://arxiv.org/abs/1701.07427} {arXiv:1701.07427 [hep-ph]} \BibitemShut
  {NoStop}%
\bibitem [{\citenamefont {Kahlhoefer}\ \emph {et~al.}(2017)\citenamefont
  {Kahlhoefer}, \citenamefont {Schmidt-Hoberg},\ and\ \citenamefont
  {Wild}}]{Kahlhoefer:2017umn}%
  \BibitemOpen
  \bibfield  {author} {\bibinfo {author} {\bibfnamefont {F.}~\bibnamefont
  {Kahlhoefer}}, \bibinfo {author} {\bibfnamefont {K.}~\bibnamefont
  {Schmidt-Hoberg}}, \ and\ \bibinfo {author} {\bibfnamefont {S.}~\bibnamefont
  {Wild}},\ }\href {\doibase 10.1088/1475-7516/2017/08/003} {\bibfield
  {journal} {\bibinfo  {journal} {JCAP}\ }\textbf {\bibinfo {volume} {1708}},\
  \bibinfo {pages} {003} (\bibinfo {year} {2017})},\ \Eprint
  {http://arxiv.org/abs/1704.02149} {arXiv:1704.02149 [hep-ph]} \BibitemShut
  {NoStop}%
\bibitem [{\citenamefont {Batell}\ \emph {et~al.}(2011)\citenamefont {Batell},
  \citenamefont {Pospelov},\ and\ \citenamefont {Ritz}}]{Batell:2009jf}%
  \BibitemOpen
  \bibfield  {author} {\bibinfo {author} {\bibfnamefont {B.}~\bibnamefont
  {Batell}}, \bibinfo {author} {\bibfnamefont {M.}~\bibnamefont {Pospelov}}, \
  and\ \bibinfo {author} {\bibfnamefont {A.}~\bibnamefont {Ritz}},\ }\href
  {\doibase 10.1103/PhysRevD.83.054005} {\bibfield  {journal} {\bibinfo
  {journal} {Phys.Rev.}\ }\textbf {\bibinfo {volume} {D83}},\ \bibinfo {pages}
  {054005} (\bibinfo {year} {2011})},\ \Eprint {http://arxiv.org/abs/0911.4938}
  {arXiv:0911.4938 [hep-ph]} \BibitemShut {NoStop}%
\bibitem [{\citenamefont {Dolan}\ \emph {et~al.}(2015)\citenamefont {Dolan},
  \citenamefont {Kahlhoefer}, \citenamefont {McCabe},\ and\ \citenamefont
  {Schmidt-Hoberg}}]{Dolan:2014ska}%
  \BibitemOpen
  \bibfield  {author} {\bibinfo {author} {\bibfnamefont {M.~J.}\ \bibnamefont
  {Dolan}}, \bibinfo {author} {\bibfnamefont {F.}~\bibnamefont {Kahlhoefer}},
  \bibinfo {author} {\bibfnamefont {C.}~\bibnamefont {McCabe}}, \ and\ \bibinfo
  {author} {\bibfnamefont {K.}~\bibnamefont {Schmidt-Hoberg}},\ }\href
  {\doibase 10.1007/JHEP07(2015)103, 10.1007/JHEP03(2015)171} {\bibfield
  {journal} {\bibinfo  {journal} {JHEP}\ }\textbf {\bibinfo {volume} {03}},\
  \bibinfo {pages} {171} (\bibinfo {year} {2015})},\ \bibinfo {note} {[Erratum:
  JHEP07,103(2015)]},\ \Eprint {http://arxiv.org/abs/1412.5174}
  {arXiv:1412.5174 [hep-ph]} \BibitemShut {NoStop}%
\bibitem [{\citenamefont {Hiller}(2004)}]{Hiller:2004ii}%
  \BibitemOpen
  \bibfield  {author} {\bibinfo {author} {\bibfnamefont {G.}~\bibnamefont
  {Hiller}},\ }\href {\doibase 10.1103/PhysRevD.70.034018} {\bibfield
  {journal} {\bibinfo  {journal} {Phys.Rev.}\ }\textbf {\bibinfo {volume}
  {D70}},\ \bibinfo {pages} {034018} (\bibinfo {year} {2004})},\ \Eprint
  {http://arxiv.org/abs/hep-ph/0404220} {arXiv:hep-ph/0404220 [hep-ph]}
  \BibitemShut {NoStop}%
\bibitem [{\citenamefont {Hahn}\ \emph {et~al.}(2010)\citenamefont {Hahn} \emph
  {et~al.}}]{Hahn:1404985}%
  \BibitemOpen
  \bibfield  {author} {\bibinfo {author} {\bibfnamefont {F.}~\bibnamefont
  {Hahn}} \emph {et~al.} (\bibinfo {collaboration} {NA62 Collaboration}),\
  }\href {https://cds.cern.ch/record/1404985} {\bibfield  {journal} {\bibinfo
  {journal} {NA62-10-07}\ } (\bibinfo {year} {2010})}\BibitemShut {NoStop}%
\bibitem [{\citenamefont {D{\"o}brich}\ \emph {et~al.}(2016)\citenamefont
  {D{\"o}brich}, \citenamefont {Jaeckel}, \citenamefont {Kahlhoefer},
  \citenamefont {Ringwald},\ and\ \citenamefont
  {Schmidt-Hoberg}}]{Dobrich:2015jyk}%
  \BibitemOpen
  \bibfield  {author} {\bibinfo {author} {\bibfnamefont {B.}~\bibnamefont
  {D{\"o}brich}}, \bibinfo {author} {\bibfnamefont {J.}~\bibnamefont
  {Jaeckel}}, \bibinfo {author} {\bibfnamefont {F.}~\bibnamefont {Kahlhoefer}},
  \bibinfo {author} {\bibfnamefont {A.}~\bibnamefont {Ringwald}}, \ and\
  \bibinfo {author} {\bibfnamefont {K.}~\bibnamefont {Schmidt-Hoberg}},\ }\href
  {\doibase 10.1007/JHEP02(2016)018} {\bibfield  {journal} {\bibinfo  {journal}
  {JHEP}\ }\textbf {\bibinfo {volume} {02}},\ \bibinfo {pages} {018} (\bibinfo
  {year} {2016})},\ \Eprint {http://arxiv.org/abs/1512.03069} {arXiv:1512.03069
  [hep-ph]} \BibitemShut {NoStop}%
\bibitem [{\citenamefont {Aidala}\ \emph {et~al.}(2017)\citenamefont {Aidala}
  \emph {et~al.}}]{Aidala:2017ofy}%
  \BibitemOpen
  \bibfield  {author} {\bibinfo {author} {\bibfnamefont {C.~A.}\ \bibnamefont
  {Aidala}} \emph {et~al.} (\bibinfo {collaboration} {SeaQuest
  Collaboration}),\ }\href@noop {} {\  (\bibinfo {year} {2017})},\ \Eprint
  {http://arxiv.org/abs/1706.09990} {arXiv:1706.09990 [physics.ins-det]}
  \BibitemShut {NoStop}%
\bibitem [{\citenamefont {Berlin}\ \emph {et~al.}(2018)\citenamefont {Berlin},
  \citenamefont {Gori}, \citenamefont {Schuster},\ and\ \citenamefont
  {Toro}}]{Berlin:2018pwi}%
  \BibitemOpen
  \bibfield  {author} {\bibinfo {author} {\bibfnamefont {A.}~\bibnamefont
  {Berlin}}, \bibinfo {author} {\bibfnamefont {S.}~\bibnamefont {Gori}},
  \bibinfo {author} {\bibfnamefont {P.}~\bibnamefont {Schuster}}, \ and\
  \bibinfo {author} {\bibfnamefont {N.}~\bibnamefont {Toro}},\ }\href@noop {}
  {\  (\bibinfo {year} {2018})},\ \Eprint {http://arxiv.org/abs/1804.00661}
  {arXiv:1804.00661 [hep-ph]} \BibitemShut {NoStop}%
\bibitem [{\citenamefont {Alekhin}\ \emph {et~al.}(2016)\citenamefont {Alekhin}
  \emph {et~al.}}]{Alekhin:2015byh}%
  \BibitemOpen
  \bibfield  {author} {\bibinfo {author} {\bibfnamefont {S.}~\bibnamefont
  {Alekhin}} \emph {et~al.},\ }\href {\doibase 10.1088/0034-4885/79/12/124201}
  {\bibfield  {journal} {\bibinfo  {journal} {Rept. Prog. Phys.}\ }\textbf
  {\bibinfo {volume} {79}},\ \bibinfo {pages} {124201} (\bibinfo {year}
  {2016})},\ \Eprint {http://arxiv.org/abs/1504.04855} {arXiv:1504.04855
  [hep-ph]} \BibitemShut {NoStop}%
\bibitem [{\citenamefont {Anelli}\ \emph {et~al.}(2015)\citenamefont {Anelli}
  \emph {et~al.}}]{Anelli:2015pba}%
  \BibitemOpen
  \bibfield  {author} {\bibinfo {author} {\bibfnamefont {M.}~\bibnamefont
  {Anelli}} \emph {et~al.} (\bibinfo {collaboration} {SHiP Collaboration}),\
  }\href@noop {} {\  (\bibinfo {year} {2015})},\ \Eprint
  {http://arxiv.org/abs/1504.04956} {arXiv:1504.04956 [physics.ins-det]}
  \BibitemShut {NoStop}%
\bibitem [{\citenamefont {Aaij}\ \emph {et~al.}(2014)\citenamefont {Aaij} \emph
  {et~al.}}]{LHCbVELOGroup:2014uea}%
  \BibitemOpen
  \bibfield  {author} {\bibinfo {author} {\bibfnamefont {R.}~\bibnamefont
  {Aaij}} \emph {et~al.},\ }\href {\doibase 10.1088/1748-0221/9/09/P09007}
  {\bibfield  {journal} {\bibinfo  {journal} {JINST}\ }\textbf {\bibinfo
  {volume} {9}},\ \bibinfo {pages} {P09007} (\bibinfo {year} {2014})},\ \Eprint
  {http://arxiv.org/abs/1405.7808} {arXiv:1405.7808 [physics.ins-det]}
  \BibitemShut {NoStop}%
\bibitem [{\citenamefont {Feng}\ \emph
  {et~al.}(2018{\natexlab{a}})\citenamefont {Feng}, \citenamefont {Galon},
  \citenamefont {Kling},\ and\ \citenamefont {Trojanowski}}]{Feng:2017uoz}%
  \BibitemOpen
  \bibfield  {author} {\bibinfo {author} {\bibfnamefont {J.~L.}\ \bibnamefont
  {Feng}}, \bibinfo {author} {\bibfnamefont {I.}~\bibnamefont {Galon}},
  \bibinfo {author} {\bibfnamefont {F.}~\bibnamefont {Kling}}, \ and\ \bibinfo
  {author} {\bibfnamefont {S.}~\bibnamefont {Trojanowski}},\ }\href {\doibase
  10.1103/PhysRevD.97.035001} {\bibfield  {journal} {\bibinfo  {journal} {Phys.
  Rev.}\ }\textbf {\bibinfo {volume} {D97}},\ \bibinfo {pages} {035001}
  (\bibinfo {year} {2018}{\natexlab{a}})},\ \Eprint
  {http://arxiv.org/abs/1708.09389} {arXiv:1708.09389 [hep-ph]} \BibitemShut
  {NoStop}%
\bibitem [{\citenamefont {Feng}\ \emph
  {et~al.}(2018{\natexlab{b}})\citenamefont {Feng}, \citenamefont {Galon},
  \citenamefont {Kling},\ and\ \citenamefont {Trojanowski}}]{Feng:2018noy}%
  \BibitemOpen
  \bibfield  {author} {\bibinfo {author} {\bibfnamefont {J.~L.}\ \bibnamefont
  {Feng}}, \bibinfo {author} {\bibfnamefont {I.}~\bibnamefont {Galon}},
  \bibinfo {author} {\bibfnamefont {F.}~\bibnamefont {Kling}}, \ and\ \bibinfo
  {author} {\bibfnamefont {S.}~\bibnamefont {Trojanowski}},\ }\href {\doibase
  10.1103/PhysRevD.98.055021} {\bibfield  {journal} {\bibinfo  {journal} {Phys.
  Rev.}\ }\textbf {\bibinfo {volume} {D98}},\ \bibinfo {pages} {055021}
  (\bibinfo {year} {2018}{\natexlab{b}})},\ \Eprint
  {http://arxiv.org/abs/1806.02348} {arXiv:1806.02348 [hep-ph]} \BibitemShut
  {NoStop}%
\bibitem [{\citenamefont {Aaij}\ \emph {et~al.}(2015)\citenamefont {Aaij} \emph
  {et~al.}}]{Aaij:2015tna}%
  \BibitemOpen
  \bibfield  {author} {\bibinfo {author} {\bibfnamefont {R.}~\bibnamefont
  {Aaij}} \emph {et~al.} (\bibinfo {collaboration} {LHCb Collaboration}),\
  }\href {\doibase 10.1103/PhysRevLett.115.161802} {\bibfield  {journal}
  {\bibinfo  {journal} {Phys. Rev. Lett.}\ }\textbf {\bibinfo {volume} {115}},\
  \bibinfo {pages} {161802} (\bibinfo {year} {2015})},\ \Eprint
  {http://arxiv.org/abs/1508.04094} {arXiv:1508.04094 [hep-ex]} \BibitemShut
  {NoStop}%
\bibitem [{\citenamefont {Aaij}\ \emph {et~al.}(2017)\citenamefont {Aaij} \emph
  {et~al.}}]{Aaij:2016qsm}%
  \BibitemOpen
  \bibfield  {author} {\bibinfo {author} {\bibfnamefont {R.}~\bibnamefont
  {Aaij}} \emph {et~al.} (\bibinfo {collaboration} {LHCb Collaboration}),\
  }\href {\doibase 10.1103/PhysRevD.95.071101} {\bibfield  {journal} {\bibinfo
  {journal} {Phys. Rev.}\ }\textbf {\bibinfo {volume} {D95}},\ \bibinfo {pages}
  {071101} (\bibinfo {year} {2017})},\ \Eprint
  {http://arxiv.org/abs/1612.07818} {arXiv:1612.07818 [hep-ex]} \BibitemShut
  {NoStop}%
\bibitem [{\citenamefont {Domingo}(2017)}]{Domingo:2016yih}%
  \BibitemOpen
  \bibfield  {author} {\bibinfo {author} {\bibfnamefont {F.}~\bibnamefont
  {Domingo}},\ }\href {\doibase 10.1007/JHEP03(2017)052} {\bibfield  {journal}
  {\bibinfo  {journal} {JHEP}\ }\textbf {\bibinfo {volume} {03}},\ \bibinfo
  {pages} {052} (\bibinfo {year} {2017})},\ \Eprint
  {http://arxiv.org/abs/1612.06538} {arXiv:1612.06538 [hep-ph]} \BibitemShut
  {NoStop}%
\bibitem [{\citenamefont {Bobeth}\ \emph {et~al.}(2001)\citenamefont {Bobeth},
  \citenamefont {Ewerth}, \citenamefont {Kruger},\ and\ \citenamefont
  {Urban}}]{Bobeth:2001sq}%
  \BibitemOpen
  \bibfield  {author} {\bibinfo {author} {\bibfnamefont {C.}~\bibnamefont
  {Bobeth}}, \bibinfo {author} {\bibfnamefont {T.}~\bibnamefont {Ewerth}},
  \bibinfo {author} {\bibfnamefont {F.}~\bibnamefont {Kruger}}, \ and\ \bibinfo
  {author} {\bibfnamefont {J.}~\bibnamefont {Urban}},\ }\href {\doibase
  10.1103/PhysRevD.64.074014} {\bibfield  {journal} {\bibinfo  {journal}
  {Phys.Rev.}\ }\textbf {\bibinfo {volume} {D64}},\ \bibinfo {pages} {074014}
  (\bibinfo {year} {2001})},\ \Eprint {http://arxiv.org/abs/hep-ph/0104284}
  {arXiv:hep-ph/0104284 [hep-ph]} \BibitemShut {NoStop}%
\bibitem [{\citenamefont {Choi}\ \emph {et~al.}(2017)\citenamefont {Choi},
  \citenamefont {Im}, \citenamefont {Park},\ and\ \citenamefont
  {Yun}}]{Choi:2017gpf}%
  \BibitemOpen
  \bibfield  {author} {\bibinfo {author} {\bibfnamefont {K.}~\bibnamefont
  {Choi}}, \bibinfo {author} {\bibfnamefont {S.~H.}\ \bibnamefont {Im}},
  \bibinfo {author} {\bibfnamefont {C.~B.}\ \bibnamefont {Park}}, \ and\
  \bibinfo {author} {\bibfnamefont {S.}~\bibnamefont {Yun}},\ }\href {\doibase
  10.1007/JHEP11(2017)070} {\bibfield  {journal} {\bibinfo  {journal} {JHEP}\
  }\textbf {\bibinfo {volume} {11}},\ \bibinfo {pages} {070} (\bibinfo {year}
  {2017})},\ \Eprint {http://arxiv.org/abs/1708.00021} {arXiv:1708.00021
  [hep-ph]} \BibitemShut {NoStop}%
\bibitem [{\citenamefont {Ali}\ \emph {et~al.}(2000)\citenamefont {Ali},
  \citenamefont {Ball}, \citenamefont {Handoko},\ and\ \citenamefont
  {Hiller}}]{Ali:1999mm}%
  \BibitemOpen
  \bibfield  {author} {\bibinfo {author} {\bibfnamefont {A.}~\bibnamefont
  {Ali}}, \bibinfo {author} {\bibfnamefont {P.}~\bibnamefont {Ball}}, \bibinfo
  {author} {\bibfnamefont {L.}~\bibnamefont {Handoko}}, \ and\ \bibinfo
  {author} {\bibfnamefont {G.}~\bibnamefont {Hiller}},\ }\href {\doibase
  10.1103/PhysRevD.61.074024} {\bibfield  {journal} {\bibinfo  {journal}
  {Phys.Rev.}\ }\textbf {\bibinfo {volume} {D61}},\ \bibinfo {pages} {074024}
  (\bibinfo {year} {2000})},\ \Eprint {http://arxiv.org/abs/hep-ph/9910221}
  {arXiv:hep-ph/9910221 [hep-ph]} \BibitemShut {NoStop}%
\bibitem [{\citenamefont {Ball}\ and\ \citenamefont
  {Zwicky}(2005{\natexlab{a}})}]{Ball:2004ye}%
  \BibitemOpen
  \bibfield  {author} {\bibinfo {author} {\bibfnamefont {P.}~\bibnamefont
  {Ball}}\ and\ \bibinfo {author} {\bibfnamefont {R.}~\bibnamefont {Zwicky}},\
  }\href {\doibase 10.1103/PhysRevD.71.014015} {\bibfield  {journal} {\bibinfo
  {journal} {Phys.Rev.}\ }\textbf {\bibinfo {volume} {D71}},\ \bibinfo {pages}
  {014015} (\bibinfo {year} {2005}{\natexlab{a}})},\ \Eprint
  {http://arxiv.org/abs/hep-ph/0406232} {arXiv:hep-ph/0406232 [hep-ph]}
  \BibitemShut {NoStop}%
\bibitem [{\citenamefont {Ball}\ and\ \citenamefont
  {Zwicky}(2005{\natexlab{b}})}]{Ball:2004rg}%
  \BibitemOpen
  \bibfield  {author} {\bibinfo {author} {\bibfnamefont {P.}~\bibnamefont
  {Ball}}\ and\ \bibinfo {author} {\bibfnamefont {R.}~\bibnamefont {Zwicky}},\
  }\href {\doibase 10.1103/PhysRevD.71.014029} {\bibfield  {journal} {\bibinfo
  {journal} {Phys. Rev.}\ }\textbf {\bibinfo {volume} {D71}},\ \bibinfo {pages}
  {014029} (\bibinfo {year} {2005}{\natexlab{b}})},\ \Eprint
  {http://arxiv.org/abs/hep-ph/0412079} {arXiv:hep-ph/0412079 [hep-ph]}
  \BibitemShut {NoStop}%
\bibitem [{\citenamefont {Bezrukov}\ and\ \citenamefont
  {Gorbunov}(2010)}]{Bezrukov:2009yw}%
  \BibitemOpen
  \bibfield  {author} {\bibinfo {author} {\bibfnamefont {F.}~\bibnamefont
  {Bezrukov}}\ and\ \bibinfo {author} {\bibfnamefont {D.}~\bibnamefont
  {Gorbunov}},\ }\href {\doibase 10.1007/JHEP05(2010)010} {\bibfield  {journal}
  {\bibinfo  {journal} {JHEP}\ }\textbf {\bibinfo {volume} {1005}},\ \bibinfo
  {pages} {010} (\bibinfo {year} {2010})},\ \Eprint
  {http://arxiv.org/abs/0912.0390} {arXiv:0912.0390 [hep-ph]} \BibitemShut
  {NoStop}%
\bibitem [{\citenamefont {Dolan}\ \emph {et~al.}(2017)\citenamefont {Dolan},
  \citenamefont {Ferber}, \citenamefont {Hearty}, \citenamefont {Kahlhoefer},\
  and\ \citenamefont {Schmidt-Hoberg}}]{Dolan:2017osp}%
  \BibitemOpen
  \bibfield  {author} {\bibinfo {author} {\bibfnamefont {M.~J.}\ \bibnamefont
  {Dolan}}, \bibinfo {author} {\bibfnamefont {T.}~\bibnamefont {Ferber}},
  \bibinfo {author} {\bibfnamefont {C.}~\bibnamefont {Hearty}}, \bibinfo
  {author} {\bibfnamefont {F.}~\bibnamefont {Kahlhoefer}}, \ and\ \bibinfo
  {author} {\bibfnamefont {K.}~\bibnamefont {Schmidt-Hoberg}},\ }\href
  {\doibase 10.1007/JHEP12(2017)094} {\bibfield  {journal} {\bibinfo  {journal}
  {JHEP}\ }\textbf {\bibinfo {volume} {12}},\ \bibinfo {pages} {094} (\bibinfo
  {year} {2017})},\ \Eprint {http://arxiv.org/abs/1709.00009} {arXiv:1709.00009
  [hep-ph]} \BibitemShut {NoStop}%
\bibitem [{\citenamefont {Arias-Aragon}\ and\ \citenamefont
  {Merlo}(2017)}]{Arias-Aragon:2017eww}%
  \BibitemOpen
  \bibfield  {author} {\bibinfo {author} {\bibfnamefont {F.}~\bibnamefont
  {Arias-Aragon}}\ and\ \bibinfo {author} {\bibfnamefont {L.}~\bibnamefont
  {Merlo}},\ }\href {\doibase 10.1007/JHEP10(2017)168} {\bibfield  {journal}
  {\bibinfo  {journal} {JHEP}\ }\textbf {\bibinfo {volume} {10}},\ \bibinfo
  {pages} {168} (\bibinfo {year} {2017})},\ \Eprint
  {http://arxiv.org/abs/1709.07039} {arXiv:1709.07039 [hep-ph]} \BibitemShut
  {NoStop}%
\bibitem [{\citenamefont {Dijkstra}\ and\ \citenamefont
  {Ruf}(2015)}]{CERN-SHiP-NOTE-2015-009}%
  \BibitemOpen
  \bibfield  {author} {\bibinfo {author} {\bibfnamefont {H.}~\bibnamefont
  {Dijkstra}}\ and\ \bibinfo {author} {\bibfnamefont {T.}~\bibnamefont {Ruf}}
  (\bibinfo {collaboration} {{SHiP Collaboration}}),\ }\href
  {http://cds.cern.ch/record/2115534} {\bibfield  {journal} {\bibinfo
  {journal} {CERN-SHiP-NOTE-2015-009}\ } (\bibinfo {year} {2015})}\BibitemShut
  {NoStop}%
\bibitem [{\citenamefont {Lanfranchi}(2017)}]{Lanfranchi:2243034}%
  \BibitemOpen
  \bibfield  {author} {\bibinfo {author} {\bibfnamefont {G.}~\bibnamefont
  {Lanfranchi}} (\bibinfo {collaboration} {SHiP Collaboration}),\ }\href
  {https://cds.cern.ch/record/2243034} {\bibfield  {journal} {\bibinfo
  {journal} {CERN-SHiP-NOTE-2017-001}\ } (\bibinfo {year} {2017})}\BibitemShut
  {NoStop}%
\bibitem [{\citenamefont {Sj{\"o}strand}\ \emph {et~al.}(2015)\citenamefont
  {Sj{\"o}strand}, \citenamefont {Ask}, \citenamefont {Christiansen},
  \citenamefont {Corke}, \citenamefont {Desai}, \citenamefont {Ilten},
  \citenamefont {Mrenna}, \citenamefont {Prestel}, \citenamefont {Rasmussen},\
  and\ \citenamefont {Skands}}]{Sjostrand:2014zea}%
  \BibitemOpen
  \bibfield  {author} {\bibinfo {author} {\bibfnamefont {T.}~\bibnamefont
  {Sj{\"o}strand}}, \bibinfo {author} {\bibfnamefont {S.}~\bibnamefont {Ask}},
  \bibinfo {author} {\bibfnamefont {J.~R.}\ \bibnamefont {Christiansen}},
  \bibinfo {author} {\bibfnamefont {R.}~\bibnamefont {Corke}}, \bibinfo
  {author} {\bibfnamefont {N.}~\bibnamefont {Desai}}, \bibinfo {author}
  {\bibfnamefont {P.}~\bibnamefont {Ilten}}, \bibinfo {author} {\bibfnamefont
  {S.}~\bibnamefont {Mrenna}}, \bibinfo {author} {\bibfnamefont
  {S.}~\bibnamefont {Prestel}}, \bibinfo {author} {\bibfnamefont {C.~O.}\
  \bibnamefont {Rasmussen}}, \ and\ \bibinfo {author} {\bibfnamefont {P.~Z.}\
  \bibnamefont {Skands}},\ }\href {\doibase 10.1016/j.cpc.2015.01.024}
  {\bibfield  {journal} {\bibinfo  {journal} {Comput. Phys. Commun.}\ }\textbf
  {\bibinfo {volume} {191}},\ \bibinfo {pages} {159} (\bibinfo {year}
  {2015})},\ \Eprint {http://arxiv.org/abs/1410.3012} {arXiv:1410.3012
  [hep-ph]} \BibitemShut {NoStop}%
\bibitem [{\citenamefont {Nadolsky}\ \emph {et~al.}(2008)\citenamefont
  {Nadolsky}, \citenamefont {Lai}, \citenamefont {Cao}, \citenamefont {Huston},
  \citenamefont {Pumplin}, \citenamefont {Stump}, \citenamefont {Tung},\ and\
  \citenamefont {Yuan}}]{Nadolsky:2008zw}%
  \BibitemOpen
  \bibfield  {author} {\bibinfo {author} {\bibfnamefont {P.~M.}\ \bibnamefont
  {Nadolsky}}, \bibinfo {author} {\bibfnamefont {H.-L.}\ \bibnamefont {Lai}},
  \bibinfo {author} {\bibfnamefont {Q.-H.}\ \bibnamefont {Cao}}, \bibinfo
  {author} {\bibfnamefont {J.}~\bibnamefont {Huston}}, \bibinfo {author}
  {\bibfnamefont {J.}~\bibnamefont {Pumplin}}, \bibinfo {author} {\bibfnamefont
  {D.}~\bibnamefont {Stump}}, \bibinfo {author} {\bibfnamefont {W.-K.}\
  \bibnamefont {Tung}}, \ and\ \bibinfo {author} {\bibfnamefont {C.~P.}\
  \bibnamefont {Yuan}},\ }\href {\doibase 10.1103/PhysRevD.78.013004}
  {\bibfield  {journal} {\bibinfo  {journal} {Phys. Rev.}\ }\textbf {\bibinfo
  {volume} {D78}},\ \bibinfo {pages} {013004} (\bibinfo {year} {2008})},\
  \Eprint {http://arxiv.org/abs/0802.0007} {arXiv:0802.0007 [hep-ph]}
  \BibitemShut {NoStop}%
\bibitem [{\citenamefont {Lourenco}\ and\ \citenamefont
  {Wohri}(2006)}]{Lourenco:2006vw}%
  \BibitemOpen
  \bibfield  {author} {\bibinfo {author} {\bibfnamefont {C.}~\bibnamefont
  {Lourenco}}\ and\ \bibinfo {author} {\bibfnamefont {H.~K.}\ \bibnamefont
  {Wohri}},\ }\href {\doibase 10.1016/j.physrep.2006.05.005} {\bibfield
  {journal} {\bibinfo  {journal} {Phys. Rept.}\ }\textbf {\bibinfo {volume}
  {433}},\ \bibinfo {pages} {127} (\bibinfo {year} {2006})},\ \Eprint
  {http://arxiv.org/abs/hep-ph/0609101} {arXiv:hep-ph/0609101 [hep-ph]}
  \BibitemShut {NoStop}%
\bibitem [{\citenamefont {Bergsma}\ \emph {et~al.}(1985)\citenamefont {Bergsma}
  \emph {et~al.}}]{Bergsma:1985qz}%
  \BibitemOpen
  \bibfield  {author} {\bibinfo {author} {\bibfnamefont {F.}~\bibnamefont
  {Bergsma}} \emph {et~al.} (\bibinfo {collaboration} {CHARM Collaboration}),\
  }\href {\doibase 10.1016/0370-2693(85)90400-9} {\bibfield  {journal}
  {\bibinfo  {journal} {Phys.Lett.}\ }\textbf {\bibinfo {volume} {B157}},\
  \bibinfo {pages} {458} (\bibinfo {year} {1985})}\BibitemShut {NoStop}%
\bibitem [{\citenamefont {Bergsma}\ \emph {et~al.}(1983)\citenamefont {Bergsma}
  \emph {et~al.}}]{Bergsma:1983rt}%
  \BibitemOpen
  \bibfield  {author} {\bibinfo {author} {\bibfnamefont {F.}~\bibnamefont
  {Bergsma}} \emph {et~al.} (\bibinfo {collaboration} {CHARM Collaboration}),\
  }\href {\doibase 10.1016/0370-2693(83)90275-7} {\bibfield  {journal}
  {\bibinfo  {journal} {Phys. Lett.}\ }\textbf {\bibinfo {volume} {128B}},\
  \bibinfo {pages} {361} (\bibinfo {year} {1983})}\BibitemShut {NoStop}%
\bibitem [{\citenamefont {Cortina~Gil}\ \emph {et~al.}(2017)\citenamefont
  {Cortina~Gil} \emph {et~al.}}]{NA62:2017rwk}%
  \BibitemOpen
  \bibfield  {author} {\bibinfo {author} {\bibfnamefont {E.}~\bibnamefont
  {Cortina~Gil}} \emph {et~al.} (\bibinfo {collaboration} {NA62
  Collaboration}),\ }\href {\doibase 10.1088/1748-0221/12/05/P05025} {\bibfield
   {journal} {\bibinfo  {journal} {JINST}\ }\textbf {\bibinfo {volume} {12}},\
  \bibinfo {pages} {P05025} (\bibinfo {year} {2017})},\ \Eprint
  {http://arxiv.org/abs/1703.08501} {arXiv:1703.08501 [physics.ins-det]}
  \BibitemShut {NoStop}%
\bibitem [{\citenamefont {Dijkstra}\ \emph {et~al.}(2015)\citenamefont
  {Dijkstra}, \citenamefont {Ferro-Luzzi}, \citenamefont {van Herwijnen},\ and\
  \citenamefont {Ruf}}]{SHiPSpectrometer}%
  \BibitemOpen
  \bibfield  {author} {\bibinfo {author} {\bibfnamefont {H.}~\bibnamefont
  {Dijkstra}}, \bibinfo {author} {\bibfnamefont {M.}~\bibnamefont
  {Ferro-Luzzi}}, \bibinfo {author} {\bibfnamefont {E.}~\bibnamefont {van
  Herwijnen}}, \ and\ \bibinfo {author} {\bibfnamefont {T.}~\bibnamefont {Ruf}}
  (\bibinfo {collaboration} {SHiP Collaboration}),\ }\href
  {https://cds.cern.ch/record/2005715} {\bibfield  {journal} {\bibinfo
  {journal} {SHiP-PUB-2015-002}\ } (\bibinfo {year} {2015})}\BibitemShut
  {NoStop}%
\bibitem [{\citenamefont {Aaij}\ \emph {et~al.}()\citenamefont {Aaij} \emph
  {et~al.}}]{LHCbcode}%
  \BibitemOpen
  \bibfield  {author} {\bibinfo {author} {\bibfnamefont {R.}~\bibnamefont
  {Aaij}} \emph {et~al.} (\bibinfo {collaboration} {LHCb Collaboration}),\
  }\href@noop {} {}\bibinfo {note} {Supplementary Material,
  \url{http://cds.cern.ch/record/2045144/files/LHCb-PAPER-2015-036-supplementary.zip}}\BibitemShut
  {NoStop}%
\bibitem [{\citenamefont {Aaij}\ \emph {et~al.}(2013)\citenamefont {Aaij} \emph
  {et~al.}}]{Aaij:2012vr}%
  \BibitemOpen
  \bibfield  {author} {\bibinfo {author} {\bibfnamefont {R.}~\bibnamefont
  {Aaij}} \emph {et~al.} (\bibinfo {collaboration} {LHCb Collaboration}),\
  }\href {\doibase 10.1007/JHEP02(2013)105} {\bibfield  {journal} {\bibinfo
  {journal} {JHEP}\ }\textbf {\bibinfo {volume} {02}},\ \bibinfo {pages} {105}
  (\bibinfo {year} {2013})},\ \Eprint {http://arxiv.org/abs/1209.4284}
  {arXiv:1209.4284 [hep-ex]} \BibitemShut {NoStop}%
\bibitem [{\citenamefont {Branco}\ \emph {et~al.}(2012)\citenamefont {Branco},
  \citenamefont {Ferreira}, \citenamefont {Lavoura}, \citenamefont {Rebelo},
  \citenamefont {Sher},\ and\ \citenamefont {Silva}}]{Branco:2011iw}%
  \BibitemOpen
  \bibfield  {author} {\bibinfo {author} {\bibfnamefont {G.~C.}\ \bibnamefont
  {Branco}}, \bibinfo {author} {\bibfnamefont {P.~M.}\ \bibnamefont
  {Ferreira}}, \bibinfo {author} {\bibfnamefont {L.}~\bibnamefont {Lavoura}},
  \bibinfo {author} {\bibfnamefont {M.~N.}\ \bibnamefont {Rebelo}}, \bibinfo
  {author} {\bibfnamefont {M.}~\bibnamefont {Sher}}, \ and\ \bibinfo {author}
  {\bibfnamefont {J.~P.}\ \bibnamefont {Silva}},\ }\href {\doibase
  10.1016/j.physrep.2012.02.002} {\bibfield  {journal} {\bibinfo  {journal}
  {Phys. Rept.}\ }\textbf {\bibinfo {volume} {516}},\ \bibinfo {pages} {1}
  (\bibinfo {year} {2012})},\ \Eprint {http://arxiv.org/abs/1106.0034}
  {arXiv:1106.0034 [hep-ph]} \BibitemShut {NoStop}%
\bibitem [{\citenamefont {Izaguirre}\ \emph {et~al.}(2017)\citenamefont
  {Izaguirre}, \citenamefont {Lin},\ and\ \citenamefont
  {Shuve}}]{Izaguirre:2016dfi}%
  \BibitemOpen
  \bibfield  {author} {\bibinfo {author} {\bibfnamefont {E.}~\bibnamefont
  {Izaguirre}}, \bibinfo {author} {\bibfnamefont {T.}~\bibnamefont {Lin}}, \
  and\ \bibinfo {author} {\bibfnamefont {B.}~\bibnamefont {Shuve}},\ }\href
  {\doibase 10.1103/PhysRevLett.118.111802} {\bibfield  {journal} {\bibinfo
  {journal} {Phys. Rev. Lett.}\ }\textbf {\bibinfo {volume} {118}},\ \bibinfo
  {pages} {111802} (\bibinfo {year} {2017})},\ \Eprint
  {http://arxiv.org/abs/1611.09355} {arXiv:1611.09355 [hep-ph]} \BibitemShut
  {NoStop}%
\bibitem [{\citenamefont {Ellwanger}\ and\ \citenamefont
  {Moretti}(2016)}]{Ellwanger:2016wfe}%
  \BibitemOpen
  \bibfield  {author} {\bibinfo {author} {\bibfnamefont {U.}~\bibnamefont
  {Ellwanger}}\ and\ \bibinfo {author} {\bibfnamefont {S.}~\bibnamefont
  {Moretti}},\ }\href {\doibase 10.1007/JHEP11(2016)039} {\bibfield  {journal}
  {\bibinfo  {journal} {JHEP}\ }\textbf {\bibinfo {volume} {11}},\ \bibinfo
  {pages} {039} (\bibinfo {year} {2016})},\ \Eprint
  {http://arxiv.org/abs/1609.01669} {arXiv:1609.01669 [hep-ph]} \BibitemShut
  {NoStop}%
\bibitem [{\citenamefont {Aaboud}\ \emph {et~al.}(2018)\citenamefont {Aaboud}
  \emph {et~al.}}]{Aaboud:2018jbr}%
  \BibitemOpen
  \bibfield  {author} {\bibinfo {author} {\bibfnamefont {M.}~\bibnamefont
  {Aaboud}} \emph {et~al.} (\bibinfo {collaboration} {ATLAS}),\ }\href@noop {}
  {\  (\bibinfo {year} {2018})},\ \Eprint {http://arxiv.org/abs/1808.03057}
  {arXiv:1808.03057 [hep-ex]} \BibitemShut {NoStop}%
\bibitem [{\citenamefont {Sirunyan}\ \emph {et~al.}(2018)\citenamefont
  {Sirunyan} \emph {et~al.}}]{Sirunyan:2018pwn}%
  \BibitemOpen
  \bibfield  {author} {\bibinfo {author} {\bibfnamefont {A.~M.}\ \bibnamefont
  {Sirunyan}} \emph {et~al.} (\bibinfo {collaboration} {CMS}),\ }\href@noop {}
  {\  (\bibinfo {year} {2018})},\ \Eprint {http://arxiv.org/abs/1808.03078}
  {arXiv:1808.03078 [hep-ex]} \BibitemShut {NoStop}%
\bibitem [{\citenamefont {D'Ambrosio}\ \emph {et~al.}(2002)\citenamefont
  {D'Ambrosio}, \citenamefont {Giudice}, \citenamefont {Isidori},\ and\
  \citenamefont {Strumia}}]{DAmbrosio:2002vsn}%
  \BibitemOpen
  \bibfield  {author} {\bibinfo {author} {\bibfnamefont {G.}~\bibnamefont
  {D'Ambrosio}}, \bibinfo {author} {\bibfnamefont {G.~F.}\ \bibnamefont
  {Giudice}}, \bibinfo {author} {\bibfnamefont {G.}~\bibnamefont {Isidori}}, \
  and\ \bibinfo {author} {\bibfnamefont {A.}~\bibnamefont {Strumia}},\ }\href
  {\doibase 10.1016/S0550-3213(02)00836-2} {\bibfield  {journal} {\bibinfo
  {journal} {Nucl. Phys.}\ }\textbf {\bibinfo {volume} {B645}},\ \bibinfo
  {pages} {155} (\bibinfo {year} {2002})},\ \Eprint
  {http://arxiv.org/abs/hep-ph/0207036} {arXiv:hep-ph/0207036 [hep-ph]}
  \BibitemShut {NoStop}%
\bibitem [{\citenamefont {Winkler}(2018)}]{Winkler:2018qyg}%
  \BibitemOpen
  \bibfield  {author} {\bibinfo {author} {\bibfnamefont {M.~W.}\ \bibnamefont
  {Winkler}},\ }\href@noop {} {\  (\bibinfo {year} {2018})},\ \Eprint
  {http://arxiv.org/abs/1809.01876} {arXiv:1809.01876 [hep-ph]} \BibitemShut
  {NoStop}%
\bibitem [{\citenamefont {Clarke}\ \emph {et~al.}(2014)\citenamefont {Clarke},
  \citenamefont {Foot},\ and\ \citenamefont {Volkas}}]{Clarke:2013aya}%
  \BibitemOpen
  \bibfield  {author} {\bibinfo {author} {\bibfnamefont {J.~D.}\ \bibnamefont
  {Clarke}}, \bibinfo {author} {\bibfnamefont {R.}~\bibnamefont {Foot}}, \ and\
  \bibinfo {author} {\bibfnamefont {R.~R.}\ \bibnamefont {Volkas}},\ }\href
  {\doibase 10.1007/JHEP02(2014)123} {\bibfield  {journal} {\bibinfo  {journal}
  {JHEP}\ }\textbf {\bibinfo {volume} {1402}},\ \bibinfo {pages} {123}
  (\bibinfo {year} {2014})},\ \Eprint {http://arxiv.org/abs/1310.8042}
  {arXiv:1310.8042 [hep-ph]} \BibitemShut {NoStop}%
\bibitem [{\citenamefont {Dev}\ \emph {et~al.}(2017)\citenamefont {Dev},
  \citenamefont {Mohapatra},\ and\ \citenamefont {Zhang}}]{Dev:2017dui}%
  \BibitemOpen
  \bibfield  {author} {\bibinfo {author} {\bibfnamefont {P.~S.~B.}\
  \bibnamefont {Dev}}, \bibinfo {author} {\bibfnamefont {R.~N.}\ \bibnamefont
  {Mohapatra}}, \ and\ \bibinfo {author} {\bibfnamefont {Y.}~\bibnamefont
  {Zhang}},\ }\href {\doibase 10.1016/j.nuclphysb.2017.07.021} {\bibfield
  {journal} {\bibinfo  {journal} {Nucl. Phys.}\ }\textbf {\bibinfo {volume}
  {B923}},\ \bibinfo {pages} {179} (\bibinfo {year} {2017})},\ \Eprint
  {http://arxiv.org/abs/1703.02471} {arXiv:1703.02471 [hep-ph]} \BibitemShut
  {NoStop}%
\end{thebibliography}
\end{document}